\documentclass[
aps,
prd,
10pt,
twocolumn,
nofootinbib,
preprintnumbers,
superscriptaddress,
showpacs]{revtex4-2}

\usepackage{comment}
\usepackage{booktabs,makecell}
\usepackage{graphicx}
\usepackage{amsmath,amssymb}
\usepackage{amsfonts}
\usepackage{xspace} 
\usepackage[usenames]{color}
\usepackage{dcolumn}
\usepackage{bm}
\usepackage{mathrsfs}
\usepackage[colorlinks=true,citecolor=blue]{hyperref}
\usepackage[all]{hypcap} 
\usepackage[utf8]{inputenc} 
\usepackage{slashed}
\usepackage{multirow}
\usepackage{rotating}
\definecolor{orange}{rgb}{1,0.5,0}
\usepackage{tabularx}
\usepackage{siunitx}
\usepackage{xfrac}
\usepackage[normalem]{ulem}
\usepackage[dvipsnames]{xcolor}

\begin{document}

\title{Missing solutions in scalar-Gauss-Bonnet gravity}

\author{O\u{g}uzhan K. Yamak}
\email{oyamak21@ku.edu.tr}
\affiliation{Department of Physics, Ko\c{c} University, Rumelifeneri Yolu,
34450 Sar{\i}yer, \.{I}stanbul, T\"{u}rkiye}

\author{Kalin V. Staykov}\email{kstaykov@phys.uni-sofia.bg}
\affiliation{Department of Theoretical Physics, Faculty of Physics,
Sofia University ``St. Kliment Ohridski", Sofia 1164, Bulgaria}

\author{Fethi M. Ramazano\u{g}lu}
\email{framazanoglu@ku.edu.tr}
\affiliation{Department of Physics, Ko\c{c} University, Rumelifeneri Yolu,
34450 Sar{\i}yer, \.{I}stanbul, T\"{u}rkiye}

\author{Daniela D. Doneva}\email{daniela.doneva@uni-tuebingen.de}
\affiliation{Departamento de Astronom\'ia y Astrof\'isica, Universitat de Val\`encia,
Av. Vicent Andrés Estellés, 19, 46100, Burjassot (Val\`encia), Spain}
\affiliation{Theoretical Astrophysics, Eberhard Karls University of T\"ubingen, 72076 T\"ubingen, Germany}

\author{Stoytcho S. Yazadjiev}\email{yazad@phys.uni-sofia.bg}
\affiliation{Department of Theoretical Physics, Sofia University ``St. Kliment Ohridski", Sofia 1164, Bulgaria}
\affiliation{Institute of Mathematics and Informatics, Bulgarian Academy of Sciences, Acad. G. Bonchev St. 8, Sofia 1113, Bulgaria}

\begin{abstract}
We discover novel neutron star solutions under scalar-Gauss-Bonnet (sGB) gravity which have been missed altogether in the existing literature. The complete solution space also reveals a new regularity condition that has to be satisfied by all scalarized stars. Certain scalarized branches of neutron star solutions under sGB gravity have long been known to abruptly terminate on stellar mass-radius diagrams at lower stellar radii, a phenomenon that has no counterpart in classical scalar-tensor theories. Such termination has been strongly suspected to be physical rather than a numerical artifact nevertheless, since sGB theories feature a regularity condition at the center of the star, which can prevent the existence of certain configurations. Motivated by the recently re-emerging  phase-transition view of scalarization, we first predict that these terminated branches can reappear after a ``gap.'' We confirm this by finding such reappearing branches, and also determine which parts of the theory parameter space are more amenable to such occurrences. We then show that the previously known central regularity condition cannot explain all the non-existing solutions in the gaps, which leads us to the discovery of a new, non-central, condition for the existence of scalarized neutron stars under sGB gravity.

\end{abstract}

\maketitle

\section{Introduction}
Neutron stars, with their large compactness and extreme core densities, provide an important testbed for fundamental physics. Despite the uncertainties in the underlying equation of state, and the consequently more involved phenomenology compared to black holes, they still provide some of the best constraints on strong field gravity~\cite{Zhao:2022vig, Silva:2020acr,LIGOScientific:2018dkp}. The presence of matter can trigger deviations from General Relativity (GR)~\cite{Damour:1992we,Damour:1993hw,Damour:1996ke,Doneva:2022ewd} that are otherwise suppressed for other compact objects, such as vacuum black holes~\cite{Yazadjiev:2025ezx}.

Neutron stars in theories beyond GR have been studied for decades. These include for example the classical scalar-tensor and $f(R)$ theories~\cite{Damour:1993hw,Damour:1996ke,Harada:1998ge,Ramazanoglu:2016kul,Yazadjiev:2016pcb,Yazadjiev:2014cza}, different flavors of scalar-Gauss-Bonnet (sGB) theory~\cite{Pani:2011xm,Doneva:2017duq,Silva:2017uqg,Xu:2021kfh,Charmousis:2021npl,Ventagli:2021ubn}, dynamical Chern-Simons gravity~\cite{Yagi:2013mbt}, and tensor-multi-scalar theories~\cite{Horbatsch:2015bua,Doneva:2019ltb}. Such investigations have included increasingly more detailed aspects of neutron star physics such as rapid rotatation~\cite{Doneva:2013qva,Yazadjiev:2015zia,Kleihaus:2016dui} and magnetic fields~\cite{Soldateschi:2020hju}. Nevertheless, the theoretically or computationally more complicated cases still remain largely unexplored and may lead to interesting new discoveries even in relatively conventional extensions of GR, such as scalar-tensor theories. 

A particularly interesting example of neutron stars under modified gravity is provided by neutron stars undergoing a so-called first-order gravitational phase transition. Phase transitions have been considered in scalar-tensor theories, spontaneous scalarization being a well-known phenomenon where certain strongly gravitating systems develop scalar hair in analogy to spontaneous magnetization~\cite{Damour:1992we,Damour:1993hw,Damour:1996ke,Doneva:2022ewd}. Nevertheless, the case has been almost exclusively considered to be a second-order (continuous) phase transition until recently.

The key idea behind first-order gravitational phase transition, specifically scalarization, is the existence of disconnected stable branches describing two phases of a neutron star: typically a bald GR-like phase and a scalarized one. First discovered in the context of maximum-mass models~\cite{Kuan:2022oxs}, it was later demonstrated that first-order gravitational phase transitions are much more common than originally thought~\cite{Doneva:2023kkz,Unluturk:2025zie,Muniz:2025egq,Ozinan:2026bne,Ozinan:2026ffv,Staykov:2026ojk}. Similar first-order phase transitions have also been found for black holes~\cite{Stefanov:2007eq,Doneva:2021tvn,Doneva:2022byd,Herdeiro:2026sur}.

The analogy between conventional matter phase transitions, for example between normal nuclear matter and quark matter, and gravitational phase transitions has turned out to be closer than one might initially expect. In the former case, a strong first-order phase transition can lead to two disconnected branches of solutions, the so-called twin stars~\cite{Glendenning:1998ag}. Similarly, gravitational first-order phase transitions appear to be a generic feature of different classes of modified theories of gravity~\cite{Kuan:2022oxs,Doneva:2023kkz,Unluturk:2025zie,Muniz:2025egq}. Moreover, their existence and qualitative behavior can be successfully described phenomenologically within Landau theory~\cite{Unluturk:2025zie}, originally developed for phase transitions in ordinary matter.

Following the discovery of first-order gravitational phase transitions, it is natural to explore the behavior of the corresponding solutions in different setups and different theories of gravity in more detail. The role and prominence of both first- and second-order phase transition in the original scalarization model of  Damour and Esposito-Far{\`e}se (DEF) are now well-understood~\cite{Kuan:2022oxs,Unluturk:2025zie,Ozinan:2026bne,Staykov:2026ojk}. Another interesting theory, where both black holes and neutron stars can scalarize is sGB gravity, whose aspects have been explored in various studies~\cite{Pani:2011xm,Doneva:2017duq,Kleihaus:2016dui},see \textcite{Doneva:2022ewd} for a review. Isolated examples of first-order phase transitions have also been reported in this theory~\cite{Doneva:2023kkz,Herdeiro:2026sur}, but most details are still unexplored.

One feature of sGB gravity that strongly differentiates it from the original DEF model of scalarization is the abrupt termination of scalarized branches where standard numerical methods cannot find any solutions beyond a given central energy density. This means, a large part of the scalarized phase, and in some cases all potentially stable scalarized solutions, are missing, which has drastic consequences for the astrophysical realization of scalarization and the signals from scalarized neutron stars. Hydrodynamical equilibrium equations under sGB gravity are known to lead to regularity conditions that can prevent neutron star configurations with arbitrarily large central energy densities~\cite{Pani:2011xm,Kleihaus:2016dui,Doneva:2017duq}, hence, the missing solutions have long been considered to be a physical feature of the theory rather than a shortcoming of the numerical computations.

This study investigates two important issues overlooked in the existing understanding of the sGB theories. First, is it really the case that the branch termination and the resulting missing solutions above are due to the regularity condition we mentioned? This was investigated for some isolated parameter values, but never in detail. Second, even if we are confident that a branch is terminated physically due to a regularity condition (as opposed to by a numerical problem), could it be possible that it reappears later on in the phase diagram? Could it be that we have overlooked such reappearance simply by not searching hard enough?

We find both questions to have surprising answers, revealing aspects of sGB gravity that were completely missed so far. For the first question, we find that the well-known regularity condition cannot explain all missing solutions. At first, this might lead to a suspicion that some of the branch terminations are simply failing numerics, but we find the case to be more surprising and illuminative: even though the known regularity condition is not violated at the termination point in some cases, we discover that there is a separate, novel condition which is violated. This second condition is related to the regularity of the solutions at some nonzero radius of the star.

We answer the second question in the affirmative, and show that the terminated branches can reappear to form, what we call, \emph{gapped} branches, which have also been completely missed in the existing literature. Two main factors played important roles in this discovery: following the lessons from a phase-transition understanding of scalarization~\cite{Unluturk:2025zie,Ozinan:2026bne, Staykov:2026ojk}, which suggested a possible continuation of the branches that terminate early on, and a comprehensive computational search of the solution space. 

These findings complete some of the missing parts of the scalarization landscape, a detailed knowledge of which is essential to using observations to test scalar-tensor theories. As such, we expect our newly discovered scalarized solutions to guide future studies on the topic.

Sec.~II starts with presenting the reduced field and hydrostatic equations in sGB gravity and continues to discuss the previously known scalarized solutions, including their classification as first-order phase transitions where applicable. Sec.~III presents the bulk of our new findings, studying the solution branch termination and reappearance. We demonstrate how previously missed neutron star solutions can be discovered via a combination of the phase transition view of scalarization, a thorough search of the parameter space and a careful study of existence conditions that are peculiar to neutron stars under sGB gravity, one of which is introduced for the first time. We provide a summary and discuss our results in Sec.~IV. The appendices provides some further results about the aforementioned existence conditions.

We use the geometric units $G=1=c$ unless stated otherwise.

\section{Scalar-Gauss-Bonnet gravity as a phase transition}
\subsection{Essentials of Scalar-Gauss-Bonnet gravity}
\label{sec: sGB gravity}
sGB gravity in the presence of matter is defined by the action
\begin{eqnarray}
S=&&\frac{1}{16\pi}\int d^4x \sqrt{-g} 
\Big[R - 2\nabla_\mu \varphi \nabla^\mu \varphi  - \nonumber\\ && V(\varphi)
+ \bar{\lambda}^2 f(\varphi){\cal R}^2_{GB} \Big] + S_{\rm matter}  (g_{\mu\nu},\chi) .\label{eq:quadratic}
\end{eqnarray}
Here $\nabla_{\mu}$ and $R$ denote the covariant derivative and the Ricci scalar with respect to the spacetime metric $g_{\mu\nu}$, respectively. $V(\varphi)$ is the potential and  $f(\varphi)$ is the coupling function for the scalar field $\varphi$.  The Gauss-Bonnet coupling constant ${\bar\lambda}$ has  dimension of $length$ and ${\cal R}^2_{GB}$ denotes the Gauss-Bonnet invariant defined by ${\cal R}^2_{GB}=R^2 - 4 R_{\mu\nu} R^{\mu\nu} + R_{\mu\nu\alpha\beta}R^{\mu\nu\alpha\beta}$ where $R_{\mu\nu}$ and $R_{\mu\nu\alpha\beta}$ are  the Ricci tensor and  the Riemann tensor respectively. $S_{\rm matter}$ is the matter action where the minimally coupled matter fields are collectively denoted by $\chi$. 

In order to find neutron star structures that can potentially carry scalar field clouds, we vary the action~\eqref{eq:quadratic} to obtain the field equations, and solve them for a static configuration that contains nuclear matter.

We consider a perfect fluid with $T^{\rm matter}_{\mu\nu}=(\rho + p)u_{\mu}u_{\nu} + pg_{\mu\nu}$ 
where $\rho$, $p$ and $u^{\mu}$ are the energy density, pressure and 4-velocity of the fluid, respectively. We adopt a standard ansatz for a static and spherically  symmetric metric  
\begin{eqnarray}
\label{eq:metric}
ds^2= - e^{2\Phi(r)}dt^2 + e^{2\Lambda(r)} dr^2 + r^2 (d\theta^2 + \sin^2\theta d\phi^2 ).
\end{eqnarray}   
We restrict ourselves to the case with $V(\varphi) = 0$, but the effects of, for example, adding a scalar mass term is well understood and it not expected to lead to significant qualitative differences~\cite{Ramazanoglu:2016kul,Ozinan:2026bne,Staykov:2026ojk}.

Finally, inserting the matter and metric assumptions into the field equations leads to a set of coupled ordinary differential equations in the radial coordinate $r$~\cite{Doneva:2017duq}
\begin{widetext}
\begin{eqnarray} 
    &&\frac{2}{r}\bigg(1+\frac{2}{r}(1-3e^{-2\Lambda})\bar{\lambda}^2\frac{df}{d\varphi}\frac{d\varphi}{dr}\bigg)\frac{d\Lambda}{dr} +\frac{e^{2\Lambda}-1}{r^2}- \frac{4\bar{\lambda}^2}{r^2}(1-e^{-2\Lambda})\bigg[\frac{df}{d\varphi}\frac{d^2\varphi}{dr^2}+(\frac{d\varphi}{dr})^2\frac{d^2f}{d\varphi^2}\bigg]-(\frac{d\varphi}{dr})^2 =
    8\pi\rho e^{2\Lambda} \label{eq: TOV_L}\\
    &&\frac{2}{r}\bigg(1+\frac{2}{r}(1-3e^{-2\Lambda})\bar{\lambda}^2\frac{df}{d\varphi}\frac{d\varphi}{dr}\bigg)\frac{d\Phi}{dr}-\frac{e^{2\Lambda}-1}{r^2}-(\frac{d\varphi}{dr})^2=8\pi pe^{2\Lambda} \label{eq: TOV_Phi}\\ 
    &&\frac{d^2\Phi}{dr^2}+\Big(\frac{d\Phi}{dr}+\frac{1}{r}\Big)\Big(\frac{d\Phi}{dr}-\frac{d\Lambda}{dr}\Big)+\Big(\frac{d\varphi}{dr}\Big)^2-\frac{4\bar{\lambda}^2}{r}e^{-2\Lambda}\bigg[\frac{d}{dr}\Big(\frac{d\Phi}{dr}\frac{df}{d\varphi}\frac{d\varphi}{dr}\Big)+\frac{d\Phi}{dr}\Big(\frac{d\Phi}{dr}-3\frac{d\Lambda}{dr}\Big)\frac{df}{d\varphi}\frac{d\varphi}{dr}\bigg]=8\pi pe^{2\Lambda} \label{eq: TOV_Phi2}\\
    &&\frac{d^2\varphi}{dr^2}+\bigg(\frac{d\Phi}{dr}-\frac{d\Lambda}{dr}+\frac{2}{r}\bigg)\frac{d\varphi}{dr}-\frac{2\bar{\lambda}^2}{r^2}\frac{df}{d\varphi}\bigg[(1-e^{-2\Lambda})\bigg(\frac{d^2\Phi}{dr^2}+\frac{d\Phi}{dr}(\frac{d\Phi}{dr}-\frac{d\Lambda}{dr})\bigg)+2e^{-2\Lambda}\frac{d\Phi}{dr}\frac{d\Lambda}{dr}\bigg]=0 \label{eq: TOV_phi}\\
    &&\frac{dp}{dr} = -(\rho+p)\frac{d\Phi}{dr} \label{eq: TOV_p}.
\end{eqnarray}
\end{widetext}

This is a modified version of the Tolman-Oppenheimer-Volkoff equations of GR, and should be supplemented with an equation of state (EoS) and proper boundary conditions to obtain the scalarized neutron star solutions. The boundary conditions impose regularity at the center of the star, 
\begin{equation}
	\left.\Lambda\right|_{r\rightarrow 0}\rightarrow 0,\quad \left. \frac{d\Phi}{dr}\right|_{r\rightarrow 0}\rightarrow 0, \quad \left.\frac{d\varphi}{dr}\right|_{r\rightarrow 0}\rightarrow 0,
\end{equation}
and asymptotic flatness at spacial infinity,
\begin{equation}
    \label{eq: asymptotic slatness}
	\left.\Lambda\right|_{r\rightarrow \infty}\rightarrow 0,\quad \left.\Phi\right|_{r\rightarrow \infty}\rightarrow 0, \quad \left.\varphi\right|_{r\rightarrow \infty}\rightarrow 0 .
\end{equation}

As explained in detail in \textcite{Doneva:2017duq}, the regularity at the stellar center can prevent the existence of a solution, especially when the energy density becomes high (also see \textcite{Pani:2011xm}). To derive this regularity condition the metric functions, the scalar field as well as the energy density and pressure are substituted with their series expansions around $r=0$, that is
\begin{equation}
    \mathcal{F}(r)=\mathcal{F}_0 + \mathcal{F}_1 r + \mathcal{F}_2 r^2 + ...
\end{equation}
for any function $\mathcal{F}$. Comparing the coefficients of different powers of $r$ then imposes constraints on the coefficients of the Taylor series expansions. It is straightforward to conclude that $\Lambda_0=0$, $\Lambda_1=0$, $\Phi_1=0$, and $\varphi_1=0$, together with the less trivial condition
\begin{eqnarray}\label{eq: central condition}
&&9  \bar{\lambda}^4  \left(\frac{df}{d\varphi}(\varphi_0)\right)^2 (\Lambda_2)^4 + 72 \pi \bar{\lambda}^4 p_0  \left(\frac{df}{d\varphi}(\varphi_0)\right)^2   (\Lambda_2)^3 - \nonumber \\ &&6\pi \rho_0  \Lambda_2 + 16 \pi^2 \rho_{0}^2= P(\Lambda_2) = 0.
\end{eqnarray}

This is an algebraic equation, a fourth-order polynomial for $\Lambda_2$, that depends on the values of the pressure $p_0$ and energy density $\rho_0$ at the center of the star, as well as the central value of the scalar field $\varphi_0$. In case no real root and thus a solution for $\Lambda_2$ exists, regular scalarized neutron stars can not be constructed. We will refer to this equation as the \emph{central existence condition,} as it is imposed at $r=0$. Since the subscript $0$ refers to the value of a variable at the center, we will use it interchangeably with a subscript c, e.g. $p_{\rm c} \equiv p_0$. 

Note that there are different ways of expressing the regularity of solutions at the center of the star. For example, there is another condition in the literature, an inequality, which holds for dilation-Gauss-Bonnet theories~\cite{Pani:2011xm}. We present how our condition~\eqref{eq: central condition} leads to the other version in Appendix~\ref{app: central_existence other form}, and will exclusively use Eq.~\eqref{eq: central condition} in this study.

Not all solutions of the above equations are scalarized. Spontaneous scalarization occurs due to an infrared mode of the scalar field assuming a tachyonic character, i.e. a negative effective squared mass~\cite{Doneva:2017bvd}. In terms of the coupling function, this means that $f(\varphi)$ should satisfy $df(\varphi)/d\varphi|_{\varphi=0}=0$, which allows the trivial case $\varphi(r) = 0$ to be a solution of the field equations. If we consider perturbations of the  scalar field on the $\varphi = 0$ background, we obtain the standard Klein-Gordon equation
\begin{equation}
    \left( \nabla_\alpha\nabla^\alpha - m^2_{\rm eff} \right)\delta \varphi= 0\, ,
\end{equation}
where the square of the effective scalar field mass $m^2_{\rm eff}$ is defined as
\begin{equation}\label{eq: meff}
    m^2_{\rm eff} = - \frac{\bar{\lambda}^2}{4}\left.\frac{d^2 f(\varphi)}{d\varphi^2 }\right|_{\varphi=0} {\cal R}^2_{GB}\,.
\end{equation}
For neutron stars the Gauss-Bonnet invariant ${\cal R}^2_{GB}$ can change its sign, therefore, if $d^2f(\varphi)/d\varphi^2|_{\varphi=0}\ne 0$ the square of the effective scalar field mass can become negative and scalar hair can be developed~\cite{Doneva:2022ewd}. Note that for static black holes only a positive sign of the second derivative leads to a tachyonic instability and scalar field development~\cite{Doneva:2017bvd,Silva:2017uqg,Antoniou:2017acq}.

Our goal in the present paper is extending on the known results in the literature and searching for nontrivial forms of phase transitions. In order to achieve that we adopt the widely used coupling function
\begin{equation}\label{eq:coupling_f1}
    f_\epsilon(\varphi) = \frac{\epsilon}{2\beta}(1- e^{-\beta(\varphi^2 + \kappa\varphi^4)})
\end{equation}
for $\epsilon=\pm 1$, which is known to lead to strong scalarization of neutron stars. The $\kappa=0$ case has been a standard coupling function in the literature since the discovery of scalarization in sGB gravity~\cite{Doneva:2017bvd}. We will discuss the role of $\kappa$ in relation to the behavior of scalarized neutron stars in Sec.~\ref{sec: sGB phase transition}. 

\begin{figure}
    \centering
    \includegraphics[width=\columnwidth]{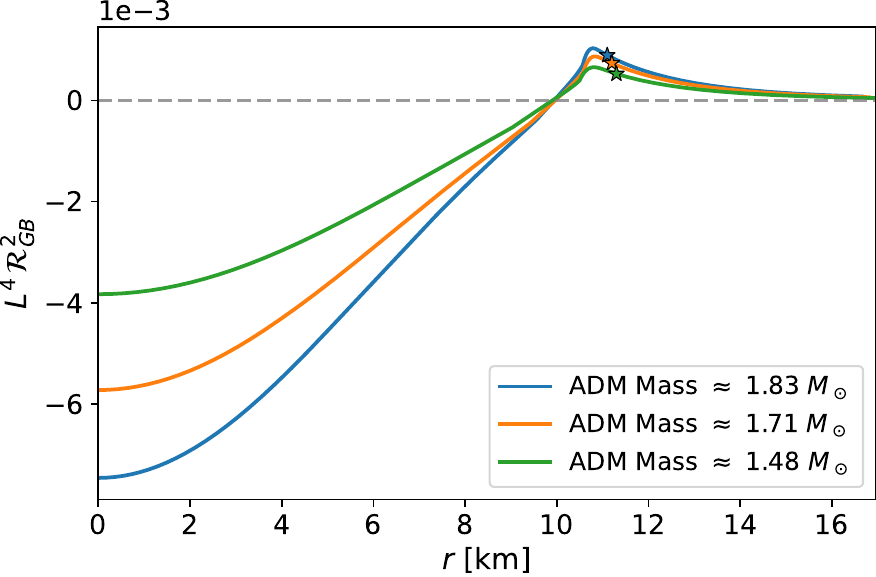}
        \caption{Gauss-Bonnet invariant as a function of $r$ for three sample neutron stars under GR. The position of the surface is marked with an asterisk in each case.}
    \label{fig: R_GB}
\end{figure}

Taking into account Eq.~\eqref{eq: meff}, let us discuss when neutron star scalarization is possible. In order to do that, we can examine the profile of ${\cal R}^2_{GB}$ for some representative neutron star solutions given in Fig.~\ref{fig: R_GB}. As evident from the figure, in most of the neutron star interior ${\cal R}^2_{GB}$ is negative and it switches sign close to the surface. Since the required condition for scalarization is $m^2_{\rm eff}<0$, we can conclude that neutron stars can develop a nontrivial scalar field for both signs of $\epsilon$~\cite{Doneva:2017duq,Silva:2017uqg}. Moreover, for $\epsilon=-1$, the scalar field is sourced by the interior of the neutron star where ${\cal R}^2_{GB}<0$, since $m^2_{\rm eff}<0$ is satisfied only in this region of the star. Thus, the scalar field acquires its maximum at the center and decreases towards the surface. The situation is reversed for $\epsilon=+1$ where the scalar field is sourced by the ${\cal R}^2_{GB}>0$ region. In this case, the scalar field has a very small value at the center, peaks close to the surface, and then decreases again away from the star. The former, $\epsilon=-1$, case has a more interesting solution structure as we will present, hence it will be the main focus of the paper. The solutions with $\epsilon=+1$ will be discussed in Sec.~\ref{sec: opposite sign}. Recalling Eq.~\eqref{eq: meff}, let us point out that only $\epsilon=+1$ leads to black hole scalarization in static configurations~\cite{Doneva:2017bvd,Silva:2017uqg,Antoniou:2017acq}.

Our results will be presented in terms of a dimensionless constant $\lambda$ defined in the following way: We fix a length scale $L = 1.47664 \, \,\mathrm{km}$ (half the solar gravitational radius) following the convention in the literature~\cite{Doneva:2017duq} and define the dimensionless $\lambda$ such that
\begin{equation*}
    \bar{\lambda}=L\lambda .
\end{equation*}
Hence, our scalarized solutions will depend on the theory parameters $(\lambda,\beta,\kappa)$. Aspects of scalarization we are interested in do not sensitively depend on the EoS, so we will use the EoS APR4~\cite{Akmal:1998cf} in its piecewise polytropic approximation form~\cite{Read:2008iy} as a representative example.

\subsection{Basics of the scalarization phase transition}
\label{sec: basic phase transition}
Spontaneous scalarization allows for the existence of both scalarized and unscalarized (GR) solutions. Transitions between these solutions can be explained by the appearance of a tachyonic mode as we explained above, but an equivalent, and sometimes more illuminative view is considering scalarization as a phase transition. For a phenomenological description, we adopt the following Landau ansatz~\cite{Unluturk:2025zie}
\begin{equation}\label{eq: landau ansatz}
    M_{\rm ADM} = M_0(M_{\rm b}) + a\eta^2 + \frac{1}{2}b\eta^4 + \frac{1}{3}c\eta^6,
\end{equation}
which expresses the total energy of the system, which is its Arnowitt-Deser-Misner (ADM) mass. Here, $M_{\rm b}$ is the baryon mass of the system that determines the number of particles that form the star, and $M_0(M_b)$ is the ADM mass of the GR star with the same baryon mass. $\eta$ is the so-called order parameter in Landau theory, which is some appropriate measure of the scalar field strength, such as the central scalar field value or scalar charge.\footnote{The order parameter appears with even powers due to the sign symmetry of the theory considered here, theories that do not possess $\varphi \rightarrow -\varphi$ symmetry can feature odd powered terms in the Landau ansatz~\cite{Staykov:2026ojk}.} 

The solutions that are realized are the ones that sit at the extrema of $M_{\rm ADM}$ as a function of $\eta$, and we have a scalarized solution when there is an extremum at $\eta_* \neq 0$. Possibility of such solutions or whether they are continuously connected to the trivial $\eta=0$ solution determine the order of the phase transition, and they are in turn determined by the coefficients $a,b,c$ which are functions of $M_{\rm b}$ as well as the EoS and the sGB theory parameters $(\lambda, \beta,\kappa)$. Thus, we can have first-order (discontinuous) or second-order (continuous) scalarization depending on these parameters. $c>0$ is assumed for the overall stability of the system. The sign of $a$ determines the stability of the unscalarized solutions ($a>0$ refers to stable and $a<0$ refers to unstable unscalarized solutions). While the sign of $b$ determines the order of the transition; $b<0$ results in a discontinuous (first-order) transition and $b>0$ results in a continuous (second-order) one.

Any form of scalarization, not just the case in sGB gravity, can be understood using the above view, but the most widely studied example which also led to the current popularity of the phase transition picture is the DEF model~\cite{Damour:1993hw}, which we now summarize.

\begin{figure}
    \centering
    \includegraphics[width=\columnwidth]{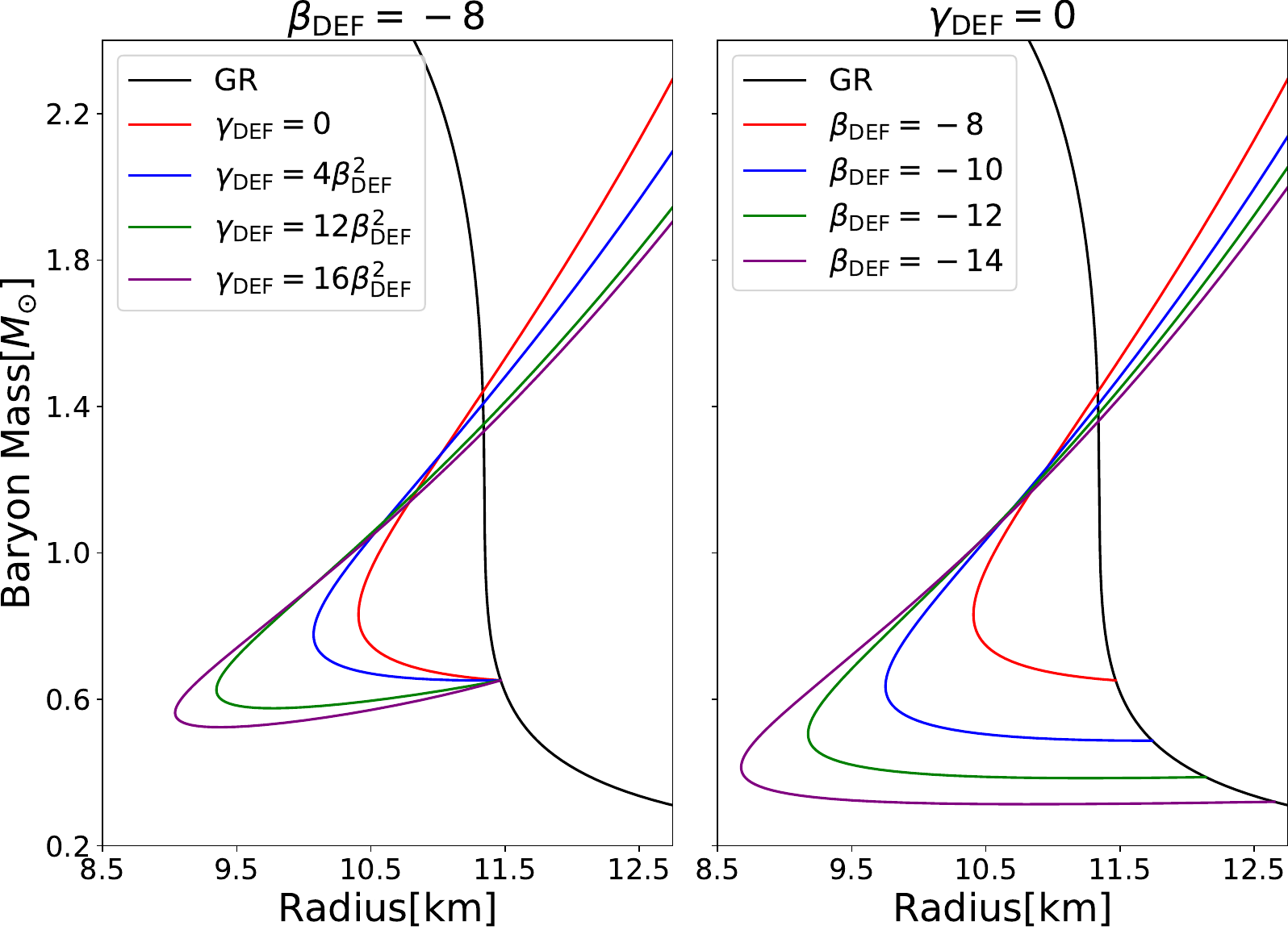}
        \caption{Phase transition picture for the DEF model as the coupling constant $\gamma_{\rm DEF}$ (left) or $\beta_{\rm DEF}$ (right) is changed, while keeping the other one constant. A positive (negative) slope, $dM/dR$, at the bifurcation point of a scalarized branch from the GR branch indicates first-order (second-order) scalarization. Increasing $\gamma_{\rm DEF}$ or $-\beta_{\rm DEF}$ facilitates first-order scalarization.}
    \label{fig: DEF phase transition}
\end{figure}
Fig.~\ref{fig: DEF phase transition} shows the solution branches in the DEF model as we change the theory parameters $(\beta_{\rm DEF},\gamma_{\rm DEF})$. Details can be found in \textcite{Ozinan:2026bne}, but they are not essential for our discussion, aside from the fact that $\beta_{\rm DEF}$ controls the quadratic (second-order in power) coupling of the scalar field, and $\gamma$ controls the quartic (fourth-order in power) coupling.\footnote{The relationship between $(\beta_{\rm DEF},\gamma_{\rm DEF})$ and the $(\lambda,\beta,\kappa)$ of Eq.~\eqref{eq:coupling_f1} is not straightforward, but in very rough terms, $-\beta$ of the DEF model behaves similarly to $\lambda$ in Eq.~\eqref{eq:coupling_f1}, and $\gamma_{\rm DEF}$ is related to all three terms in Eq.~\eqref{eq:coupling_f1}. The overall effect of increasing $\gamma_{\rm DEF}$ is qualitatively similar to increasing $\kappa$~\cite{Ozinan:2026bne}.} Note that we use the baryon mass $M_{\rm b}$ in Fig.~\ref{fig: DEF phase transition} instead of the ADM mass in Eq.~\eqref{eq: landau ansatz}, as typical in the phase transition discussions of scalarization~\cite{Unluturk:2025zie}. Both masses have the same order of magnitude and have their extremum points at the same radii, hence provide the same qualitative picture.

As we increase the quartic coupling $\gamma_{\rm DEF}$ while keeping the quadratic coupling constant, we expect the quartic $b$ term in the Landau ansatz~\eqref{eq: landau ansatz} to decrease, hence a move towards first-order phase transition. This is indeed the case in the left panel of Fig.~\ref{fig: DEF phase transition}, where the change of the slope $dM/dR$ at the bifurcation point (the point where the scalarized solutions join the GR branch) from negative to positive directly indicates the sign change for $b$. The general picture is that the part of the scalarized branch with a positive slope is unstable. The first (locally) stable scalarized solution that have a chance to be astrophysically realized, the beginning point of the solid curve at the bottom, is not continuously connected to the GR branch, but is a non-perturbatively different configuration. This is the discontinuous nature of the phase transition.

For the case of $\gamma_{\rm DEF}=0$ in the right panel of Fig.~\ref{fig: DEF phase transition}, increasing $|\beta_{\rm DEF}|$ also leads to first-order scalarization, but this also means that the discontinuous jump from the unscalarized to scalarized solutions occurs at a relatively low neutron star mass, whose astrophysical relevance can be low.

An important aspect of the scalarized solutions of the DEF model, stable or not, is that once they branch off from the GR sequence, they form a single continuous curve that can reach high neutron star masses. We will see in the next chapter that this is not always the case for the sGB theory, where the branches can terminate prematurely, soon after the bifurcation point. This will lead to a distinct phase transition picture and astrophysical phenomenology.

\subsection{Phase transition view of neutron stars in sGB gravity}
\label{sec: sGB phase transition}
The phase transition picture of black holes in sGB gravity was recently studied~\cite{Herdeiro:2026sur}, but there is no existing work for neutron stars to the best of our knowledge. The matter is far from checking yet another gravitational object, since neutron star scalarization has qualitative differences as we will present.

Let us first summarize the established behavior of neutron stars in the sGB theory before we explain them in the phase transition language. Changing $\lambda$ does not change the solution structure qualitatively, the main effect being moving the bifurcation point to different points on the GR solution curve. Thus, we will concentrate on a single reference value of $\lambda$
and investigate the effect of changing $\beta$ and $\kappa$ in Eq.~\eqref{eq:coupling_f1}.

Before we continue to the details, let us first emphasize that the $\kappa$ parameter is not introduced merely as a theoretical contrivance. Rather, it controls the quartic coupling of the scalar field to curvature, which directly affects the quartic term in the Landau ansatz~\eqref{eq: landau ansatz}. This, in turn, determines the order of the phase transition. In some sense, rather than $\kappa$ being introduced as a new term, one can argue that it is the $\kappa = 0$ case that is artificial in the phase transition view. $\kappa = 0$ is restricted since both quartic and quadratic coupling of the scalar field to gravity, and all coupling orders for that matter, are controlled by the single parameter $\beta$, for which there is no apriori reason. The specific coupling in Eq.~\ref{eq:coupling_f1} with $\kappa =0$ is mostly based on historical precedence and simplicity. Hence, it is natural to introduce a separate coupling constant at the quartic order (and not the others) since it plays a central role in scalarization as a phase transition~\cite{Ozinan:2026bne,Ozinan:2026ffv}.

\begin{figure}
    \centering
    \includegraphics[width=\columnwidth]{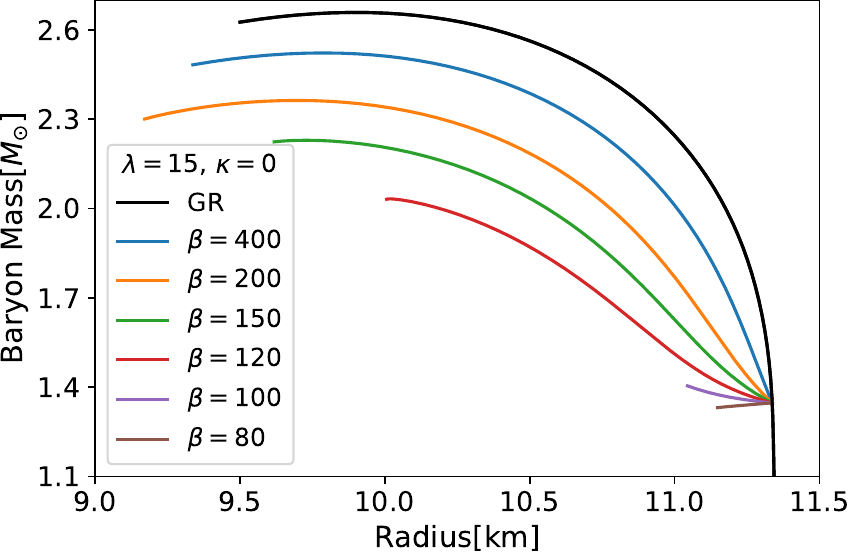}
    \caption{The scalarized branches for varying $\beta$ and $\kappa = 0$ as they are commonly presented in the literature. The order of the phase transition switches from second to first as $\beta$ decreases, but first-order scalarized branches typically terminate shortly after bifurcating from the GR branch.}
    \label{fig: M vs radius kappa=0}
\end{figure}
The solution structure, as commonly presented in the literature, for the $\kappa=0$ case is depicted in Fig.~\ref{fig: M vs radius kappa=0}. The scalarized branch is close to the GR one (but it is always below it) for high values of $\beta$, which is natural since $\beta \to \infty$ approaches the GR limit. When $\beta$ decreases, two main qualitative changes occur: the maximal mass shifts to lower masses and the branch terminates earlier. Even more prominently, the termination eventually becomes so early that the turning point of the curve where the maximal stellar mass is achieved is not reached. With further decrease in $\beta$, the slope of the solution branch also reverses, which signals instability of these solutions~\cite{Unluturk:2025zie}, and the termination occurs almost immediately after the bifurcation point.

We need to elaborate what we mean by ``termination'' of a branch. The solutions on these branches are calculated numerically, hence there is always the possibility that the computational setup is failing at these points rather than the actual non-existence of solutions. This issue will be discussed in detail in the coming sections, but it suffices for now to say that the termination is indeed physical with the central existence condition, Eq.~\eqref{eq: central condition}, being the culprit. A basic examination of the condition shows that lower $\beta$ values (for fixed $\lambda$) makes it harder for scalarized solutions to be regular at the center. Thus, it is expected that highly scalarized solutions cease to exist as $\beta$ sufficiently decreases, as we observe.

\begin{figure}
    \centering
    \includegraphics[width=\columnwidth]{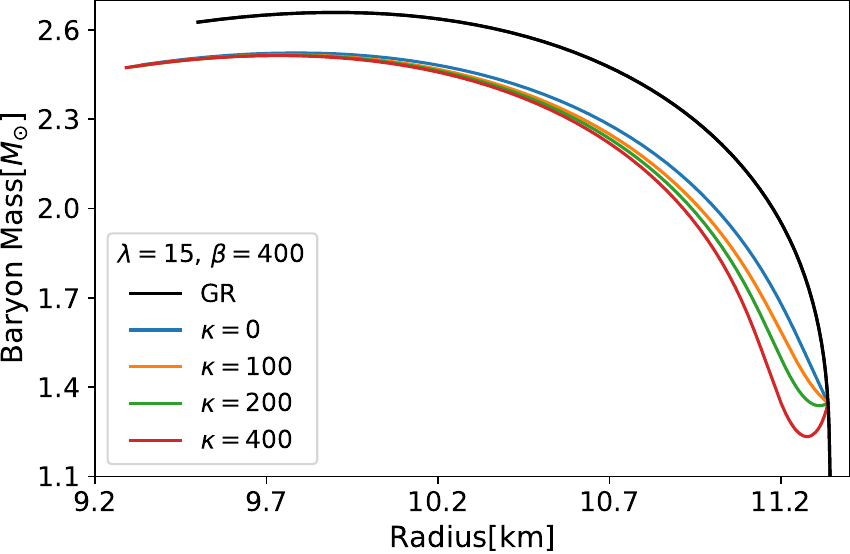}
    \caption{The scalarized branches for a fixed $\beta$ and varying $\kappa \neq 0$ as they are commonly presented in the literature. The order of the phase transition still switches from second to first simmilarly to Fig.~\ref{fig: M vs radius kappa=0}, this time as $\kappa$ increases, but there is no premature termination of the branches.}
    \label{fig: M vs radius kappa neq 0}
\end{figure}
Scalarized branches for the $\kappa \neq 0$ case are presented in Fig.~\ref{fig: M vs radius kappa neq 0}. Let us focus on a relatively high value of $\beta = 400$ which leads to a more ``complete'' branch compared to Fig.~\ref{fig: M vs radius kappa=0}, reaching a maximum mass before being terminated. Within the considered wide range of $\kappa$ values, the length of the branch does not vary in a noteworthy manner, and there are no significant deviations from the $\kappa = 0$ case regarding the maximum masses. However, there are interesting deviations in the lower stellar mass sections of the branches close to the bifurcation point. Increasing $\kappa$ eventually changes the slope of the curve at the bifurcation point, like the low $\beta$, $\kappa=0$ case in Fig.~\ref{fig: M vs radius kappa=0}, but the solution curves do not terminate early unlike before.

Now that we know the general solution structure, we can read the phase transition picture in analogy to that of the DEF model in Sec.~\ref{sec: basic phase transition}. Let us start with the $\kappa \neq 0$ example since it directly resembles the DEF model, where increasing $\kappa$ moves us towards first order scalarization similarly to the behavior with increasing $\gamma_{\rm DEF}$ in the left panel of Fig.~\ref{fig: DEF phase transition}. This is not surprising in retrospect, since both parameters are directly related to the quartic coupling of the scalar field, which strongly affects the order of the phase transition as we mentioned. This is also intimately related to the \emph{nonlinear scalarization} phenomena~\cite{Doneva:2023kkz}.

The $\kappa=0$ case in Fig.~\ref{fig: M vs radius kappa=0} has novel elements compared to the $\kappa \neq 0$ examples when we consider the parameter range in Fig.~\ref{fig: M vs radius kappa neq 0}, the most striking being the lowest $\beta$ branches. These terminate exceptionally early, which can be describes as ``failed'' first-order scalarization. Moreover, even the longer branches for intermediate $\beta$ values can terminate prematurely, before reaching an extremal stellar mass. These missing parts of the phase transition landscape can have interesting astrophysical implications. 

Recall that one of the attractive features of first order scalarization is its discontinuous nature where the transition can have violent dynamics and lead to large observable signals~\cite{Kuan:2022oxs,Ozinan:2026ffv,M:2026yut}, but there are even more peculiar features in scalarized solutions of sGB gravity. For example, for the prematurely terminated branches for $\kappa=0$ at intermediate $\beta$ values susch as $\beta = 120$ in Fig.~\ref{fig: M vs radius kappa=0}, a scalarized neutron star that increases its mass, say by accretion, can eventually reach the termination point, after which it has to find another stable branch, if there is any. In our case, the GR branch is tachyonically unstable, meaning there is no known stable solution to jump to. Whether this signals a failure of the theory to explain large mass neutron stars or the likely outcome beyond the branch termination is collapse into a black hole, is a question which requires a dedicated analysis, which is out of the scope of the present paper.

The case is even more puzzling for in what we called failed first-order scalarization, e.g. for $\kappa=0$ and $\beta=80$ in Fig.~\ref{fig: M vs radius kappa=0}. These branches terminate after a short while, and the usual stability arguments indicate, though do not prove, that the solutions on the short branches are unstable ones~\cite{Unluturk:2025zie}. Hence, once the stellar mass reaches the bifurcation point, there are no astrophysically viable stars with higher mass.

The peculiarities above both assume the fact that the numerical solutions in the literature so far are reliable, and all solutions are represented in Figs.~\ref{fig: M vs radius kappa=0} and~\ref{fig: M vs radius kappa neq 0}. As we mentioned, the central existence condition~\eqref{eq: central condition} indeed makes this likely, but this issue has not been investigated in detail so far, and some questions remain to be answered.
\begin{itemize}
    \item Are the terminal points of the scalarized branches really points where the central existence condition~\eqref{eq: central condition} ceases to have solutions? In other words, do scalarized solutions really become impossible if we go slightly beyond these points?
    
    \item Even if the terminal points are physical, is it possible that the solution branch reappears somewhere else, for example, for the failed branches at low $\beta$ in Fig.~\ref{fig: M vs radius kappa=0}? In other words, can these short branches be longer, but with a gap in the middle? 
    
    \item Similarly to the previous question, is it ever the case for $\kappa \neq 0$ that there are gaps in the branches, even though it is not the case in Fig.~\ref{fig: M vs radius kappa=0}, where we have a more standard first- and second-order transition picture?
\end{itemize}

This is not an abstract line of inquiry. Recent work demonstrated that understanding the phase transition picture of scalarization can help discover solution branches that could be missed otherwise~\cite{Staykov:2026ojk}. The central existence condition and other related constraints we will present bring a new ingredient to the phase transition understanding, which can lead to the discovery of novel structures, such as gaps on scalarized branches. These can provide novel observational possibilities, for example, a reappearing scalarized branch can provide stable configurations where an unscalarized neutron star jump into, whereas there is no such possibility, for example, in the current branches of failed first-order scalarization in Fig.~\ref{fig: M vs radius kappa=0}.

\section{Discovering missing solutions in sGB gravity}
\label{sec: solution space}
Here we perform a more thorough search of the parameter space in order to answer the questions raised in the previous section. We first present our results for $\kappa =0$ and $\kappa \neq 0$ separately, since their behaviors are distinct. Afterwards, we explain the methods that enabled us to find scalarized solutions that have been missed so far.

\subsection{$\kappa=0$: Small deviations from the established picture}
\label{sec: kappa 0}
We found that the picture in the top panel of Fig.~\ref{fig: M vs radius kappa=0} remains similar for most values of $\lambda$ and $\beta$. That is, the terminated branches do not reappear after a gap. We ensured that this is the not a numerical failure via the methods we will elaborate in Sec.~\ref{sec: nonexistence methods}.

\begin{figure}
    \centering
    \includegraphics[width=\columnwidth]{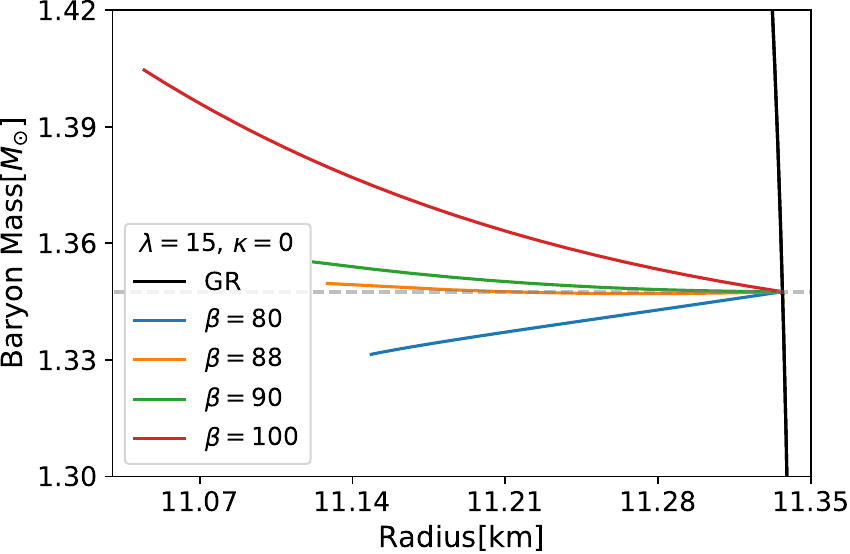}
        \caption{The scalarized branches for $\kappa=0$ and varying $\beta$, where the values of $\lambda$ and $\beta$ are adjusted such that a first-order phase transition is evident. Namely, the orange $\beta=88$ case is a scalarized branch with first-order scalarization and the branch survives long enough to change its slope. This is unlike the lowest $\beta$ curves in Fig.~\ref{fig: M vs radius kappa=0} or the $\beta=80$ curve here. This means the negative slope part of the $\beta=88$ branch, which has potentially stable configurations, might provide astrophysically viable stars, albeit in a short range of stellar masses and radii. Such branches only appear at $\beta$ values around where the phase transition order switches from second to first, hence require some fine tuning of the theory parameters.}
    \label{fig: nonfailing first order}
\end{figure}
We said the picture is similar, but there are also exceptions. When a branch has a positive slope at the bifurcation point, $dM/dR>0$, recall that it terminates pretty soon for the coupling constant values presented in Fig.~\ref{fig: M vs radius kappa=0}. Moreover, the termination occurs before the slope can change sign, which strongly indicates that all solutions on this short branch are unstable. We called this failed first-order scalarization.

As a mild deviation from the usual, this failure can be ameliorated as can be seen in Fig.~\ref{fig: nonfailing first order}, where the termination occurs after the slope of the branch reverses, leading to potentially stable scalarized solutions in first-order scalarization, albeit in a short range of stellar mass and radius values. This happens for a small window of $\beta$ values (for each $\lambda$), around where scalarization changes from second-order to first-order. As we emphasized, this is a somewhat fine-tuned case, but it is interesting since this is the first time a viable first-order scalarized solution is presented for $\kappa = 0$.

A perhaps more interesting case would be the reappearance of a branch after a gap, which can be especially interesting if the reappeared section is long enough for astrophysical relevance. This is not pure curiosity, and there is a motivation to expect such occurrences based on the behavior of scalarized branches.

First, note that the scalar field appears in the central existence condition, Eq.~\eqref{eq: central condition}, via the term
\begin{equation}\label{eq: existence coefficient}
    \lambda^4 \left[\frac{df_-}{d\varphi}(\varphi_c)\right]^2 =\lambda^4 (\varphi_{\rm c} + 2  \kappa \varphi_{\rm c}^3)^2\ e^{-2\beta(\varphi_{\rm c}^2 + \kappa \varphi_{\rm c}^4)}
\end{equation}
When $\kappa = 0$, this expression has a profile which peaks around $\varphi_{\rm c}= \frac{1}{\sqrt{2\beta}}$. In other words, the coefficients of the two highest powers in Eq.~\eqref{eq: central condition} start to diminish in the $\varphi_{\rm c} \gtrsim \frac{1}{\sqrt{2\beta}}$ regime, which makes it easier for the central existence condition to be satisfied. Second, $\varphi_{\rm c}$ tends to increase as we move away from the bifurcation point. Hence, combining these two facts indicates that a hypothetical reappearance of a terminated branch at a lower stellar radius, higher central density and higher $\varphi_{\rm c}$, becomes more probable as $\beta$ increases.

On the other hand, increasing $\beta$ in an unbouned manner weakens the overall strength of the coupling, $\beta \to \infty$ resulting in no termination at all and eventually recovering the GR limit. To be able to have termination of the branches despite large $\beta$, $\lambda$ can be increased to increase the coefficients that have the term in Eq.~\eqref{eq: existence coefficient}. Hence, for sufficiently large $\beta$ and $\lambda$, we can observe both the termination and the reappearance of scalarized branches, that is, gapped branches.

\begin{figure}
    \centering
    \includegraphics[width=\columnwidth]{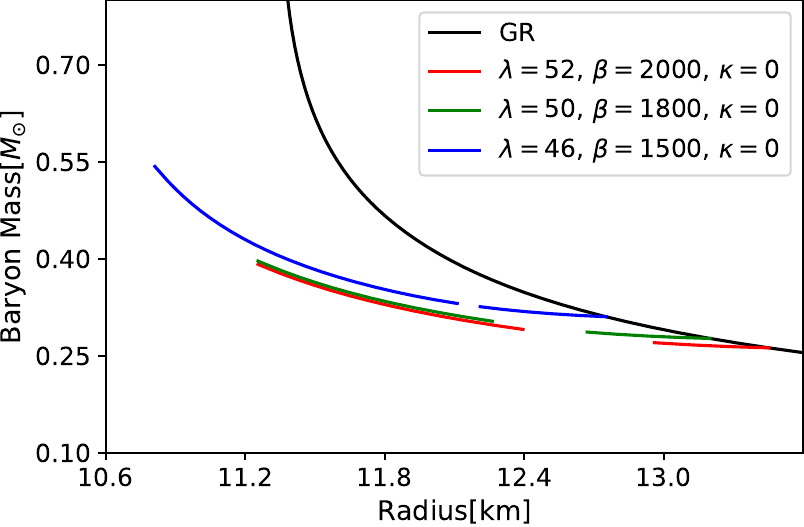}
        \caption{Representative scalarized solution branches that includes a gap for $\kappa=0$. We found such solutions to be rare, mainly encountered for higher $\lambda$ and $\beta$ values. Higher $\lambda$ means these branches bifurcate from the GR solution at very low stellar mass values, which reduces their astrophysical relevance.}
    \label{fig: gap kappa = 0}
\end{figure}
Fig.~\ref{fig: gap kappa = 0} shows some sample parameter values where a terminated short branch reappears, forming a gap between the two sections. This answers one of our main questions, whether the reappearance of terminated branches are possible, and shows that naive numerical studies of scalarization can easily miss relevant solutions. Without our targeted search in mind, the common methodology would stop looking for solutions once a scalarized branch is terminated. 

We should mention some caveats of the gapped solutions we found. For one, these clearly occur under some fine-tuning of the theory parameters, and can be argued to be exceptional cases rather than occurring on appreciably large regions of the theory parameter space. Second, the need for large $\lambda$ values typically push the gaps towards low stellar masses, which are astrophysically less interesting. The conceptual finding is nevertheless quite interesting despite these shortcomings, and we will see that the picture becomes more interesting for $\kappa \neq 0$.

\subsection{$\kappa \neq 0$: Gapped branches and other surprises}
\label{sec: kappa neq 0}
%
\begin{figure*}
    \centering
    \includegraphics[width=0.66\columnwidth]{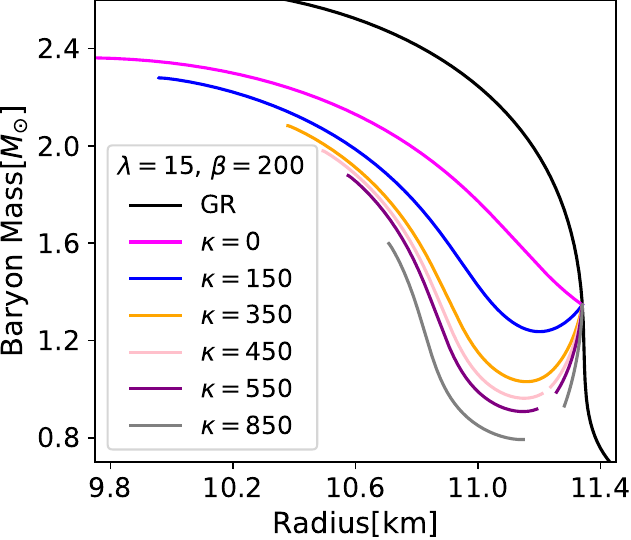}
    \includegraphics[width=0.66\columnwidth]{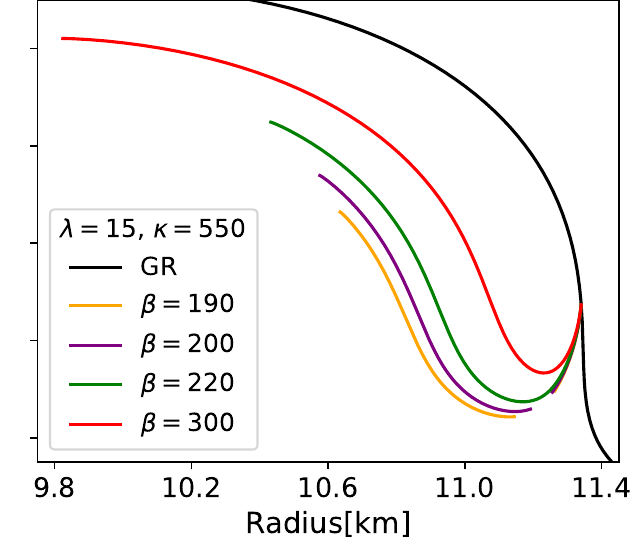}
    \includegraphics[width=0.66\columnwidth]{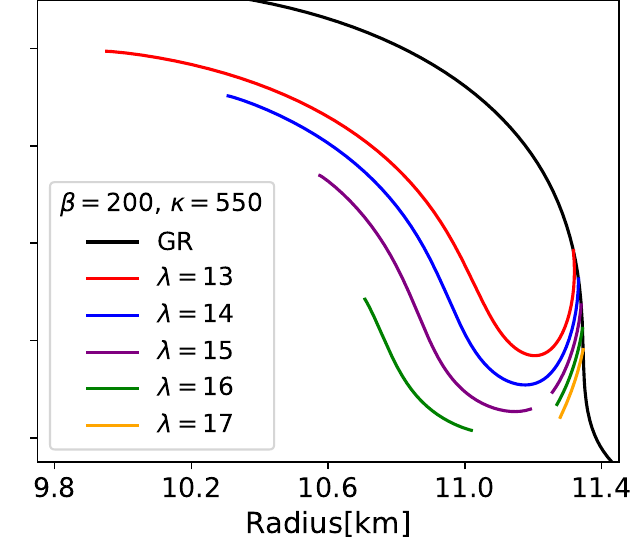}
    \label{fig: kappa neq 0 param changing}
    \caption{The behavior of the scalarized branches and the appearance of gaps as the three theory parameters $\kappa$ (left), $\beta$ (middle) and $\lambda$ (right) are changed. Decreasing $\beta$ and increasing the other two parameters facilitate branch termination and gaps. }
\end{figure*}

Recall that the specific $\kappa \neq 0$ examples in Fig.~\ref{fig: M vs radius kappa neq 0} presented a phase transition picture much closer to the basic one in Sec.~\ref{sec: basic phase transition}, in that scalarization is second order for smaller $\kappa$, and moves towards first-order as $\kappa$ increases. Moreover, there is no failed first-order scalarization where the branch terminates early on as it happens in the lower $\beta$ branches for $\kappa=0$ in Fig.~\ref{fig: M vs radius kappa=0}. A continuity argument would suggest that for $\kappa \neq 0$, decreasing $\kappa$ and $\beta$ values might lead us to failing first-order scalarization as in the low $\beta$ curves in Fig.~\ref{fig: M vs radius kappa=0}. Hence, there should not be a strict dichotomy between the $\kappa = 0$ picture of Fig.~\ref{fig: M vs radius kappa=0} and the $\kappa \neq 0$ picture of Fig.~\ref{fig: M vs radius kappa neq 0}.

This can be confirmed by checking the trend with increasing $\kappa$ as in the left panel of Fig.~\ref{fig: kappa neq 0 param changing}, where we also use a lower value of $\beta$ compared to Fig.~\ref{fig: M vs radius kappa neq 0}, whose significance will become apparent later. First of all, we see that the scalarized branches behave similarly to the $\kappa = 0$ example when $\kappa \lesssim 400$, in that they resemble second-order scalarization for lower $\kappa$, and first order for higher values, and there are no gaps on the scalarized branches. For the higher values in the $\kappa \lesssim 400$ range, the branches terminate somewhat earlier than the $\beta =400$ example in Fig.~\ref{fig: M vs radius kappa neq 0}, but they do not terminate prematurely.

Perhaps the more striking findings on this panel are for $\kappa \gtrsim 400$, which feature gapped branches we have been primarily searching for. This can be interpreted in two parts. First, the scalarized branches terminate after a short descent in stellar mass for $\kappa=450, 550, 850$. Second, however, unlike the $\kappa = 0$ case in Fig.~\ref{fig: M vs radius kappa=0}, the branch reappears at a lower radius (higher central density). The combination of these two form gapped branches.

The appearance and widening of the gaps with increasing $\kappa$ can be understood with a similar reasoning to the $\kappa = 0$ case, since the main effect of $\kappa$ is to both shift and increase the peak of the profile in Eq.~\eqref{eq: existence coefficient}. This has a total effect similar to that of increasing $\beta$ and $\lambda$ together. Hence we observe gapped branches by increasing $\kappa$ alone, while having more commonly considered values for $\beta$ and $\lambda$. A noteworthy outcome of this is having gaps at more astrophysically interesting stellar masses, $M \gtrsim 1M_\odot$, without fine tuning.

We can see the effect of $\beta$ for nonzero $\kappa$ in the middle panel of Fig.~\ref{fig: kappa neq 0 param changing}. Lower $\beta$ promotes early termination, hence provides an opportunity, but not a guarantee, for gapped branches, hence our choice of a lower $\beta$ value in the left panel. Thus, there is an interplay of $\kappa$ and $\beta$ in facilitating the gaps. However, we should add that very high $\kappa$ values can eventually prevent the reappearance of a branch altogether as well, leading to failed branches rather than gaps. 

Increasing $\lambda$ also has a similar effect on gaps to increasing $\kappa$ as seen in the right panel of Fig.~\ref{fig: kappa neq 0 param changing}. However, arguably the most prominent role of $\lambda$ is determining where the bifurcation point is, which is important for astrophysical relevance. The bifurcation point indirectly determines the stellar mass range where the gap appears. Hence, $\lambda \lesssim 15$ is favored for observability as it moves this range towards and above $1M_\odot$.

\subsection{Identifying the mechanisms for branch termination}
\label{sec: nonexistence methods}
If there is no scalarized solution where we naively expect one, for example after a termination point, we need to check two main issues in light of our previous discussion. First, we need to confirm that the existence condition~\eqref{eq: central condition} is indeed the culprit for the termination of branches. In other words, the polynomial in Eq.~\eqref{eq: central condition} smoothly changes as we move on a scalarized branch and it ceases to have any solutions for $\Lambda_2$ at the termination point. This way, we know that there is indeed a physical and mathematical reason for not finding a solution, and it is not a shortcoming of our numerical setup.

Second, even if we know that there is a physical impediment to the continuation of a branch at some point, we still need to systematically search the whole parameter space to make sure that our numerical setup did not miss a reappearance of the branch. 

In order to understand our methods to ensure that we find all the solution branches, recall how we calculate a stellar structure. To summarize, we numerically solve the modified TOV equations~\eqref{eq: TOV_L}-\eqref{eq: TOV_p}, a set of coupled ordinary differential equations together with initial conditions at the center of the star and boundary conditions at infinity. This is typically implemented by reducing the equation system~\eqref{eq: TOV_L}-\eqref{eq: TOV_p} into a first order form, which means our variable set is $U = (\Lambda, \Phi', \varphi, \varphi',  p)$. For any given initial value $p_0$, there will be only one nonzero value of $\varphi_0$ that leads to an asymptotically flat spacetime in Eq.~\eqref{eq: asymptotic slatness}, which is \emph{the} solution we seek.\footnote{There will actually be a discrete set of allowed $\varphi_0$ values in general, but all but one of them have nodes in $\varphi(r)$, and these are known to be generically unstable. We do not study such solutions as we mentioned at the end of Sec.~\ref{sec: sGB gravity}. } We find this specific $\varphi_0$ value via some search algorithm, such as the shooting method~\cite{Doneva:2017duq} or relaxation~\cite{Rosca-Mead:2020bzt,Tuna:2022qqr,Demirboga:2023ktt}. If there is no such $\varphi_0$, we conclude that there is no (scalarized) solution. To construct a whole scalarized solution branch, we slightly change $p_0$ and repeat the process to construct a series of solutions.

It is evident from our explanation that what we deduce to be a case of no solution might be a shortcoming of our search algorithm. Moreover, even if the branch terminates for physical rather than numerical reasons, it can reappear somewhere else as we saw. Then, we do not know how and where to find the point of reappearance.

To minimize chances of such shortcomings, we searched a very large portion of the $(p_0,\varphi_0)$ plane on a fine grid in a brute-force manner. This is as opposed to, for example, the shooting method, where one tries a more limited set of $\varphi_0$ values, guided by a nonlinear root-finding algorithm. The brute-force search is a very inefficient and wasteful method, but it ensures that all values are checked, and it can be manageable in specific cases such as ours. In a sense, we trade efficiency for fidelity. On one hand, we will see that all branch terminations identified with the older, usual methods were confirmed to be physical, that is, the older numerical methods were successful in this case. On the other hand, however, the brute-force search enabled us to discover the reappearance of branches.

A third point to check about our numerical solutions appeared during our brute-force searches, where some of the branch terminations could not be explained in terms of the violation of the central existence condition, Eq.~\eqref{eq: central condition}. Essentially, even though the central existence condition is satisfied, the modified TOV equations we need to integrate to construct the star could not be integrated all the way to infinity. This happens when the differential equation system becomes singular at a nonzero radius, which we dub the \emph{non-central existence condition.} 

To be more precise, the radial evolution of our variables $U(r)$ can be cast into the format
\begin{equation}
    \mathcal{M}(r)\ U' = S(U),
    \label{eq:M_radial_evolution}
\end{equation}
with $S(U)$ being the right hand side of the equation. The matrix $\mathcal{M}(r)$ should be non-singular everywhere inside the star for a solution to exist. We will see that the termination of some solution branches arise from the violation of this condition. To the best of our knowledge, this issue has not been identified in any previous work. 

To check the non-central existence condition, we integrate the field equations and calculate the determinant of $\mathcal{M}(r)$ at each point. Although the vanishing of the determinant is not strictly sufficient for non existence of a solution to the differential equation, it is still a very strong indicator. The raw determinant can strongly depend on the relative normalization of the equations and variables. Thus, we apply diagonal row and column equilibration to $\mathcal{M}(r)$ to obtain ${\cal M}_{\rm eq}(r)$,\footnote{To obtain ${\cal M}_{\rm eq}(r)$ we first divide each row by that row's euclidean norm, then we divide each column of the resulting matrix by the euclidean norm of that column. By this normalization procedure we get determinant values at the order of unity in regular cases. This is only done before checking the determinant and is not used in the radial integration.} and find the minimum value of the determinant throughout the integration region,
\begin{equation}
    \Delta_{\min}
    =\min\left|\det \mathcal{M}_{\rm eq}(r)\right|.
    \label{eq:det_diagnostic}
\end{equation}
The integration should reach the stellar surface without encountering $\Delta_{\min}=0$, which is the non-central existence condition. Hence,
\begin{equation}
    \Delta_{\min}=0
    \label{eq: noncentral existence condition}
\end{equation}
anywhere inside the star identifies a violation of our condition. Numerically, we identify the problem as $\Delta_{\min} \ll 1$, and the fact that this criterion coincides with the lack of solutions in our brute-force search further supports the idea that the problem is indeed not a numerical glitch. We also discovered a close mathematical relationship between the central and non-central existence conditions, which we elaborate on in Appendix~\ref{app: two conditions are related}. 

\begin{figure}[t]
    \centering
    \includegraphics[width=0.5\textwidth]{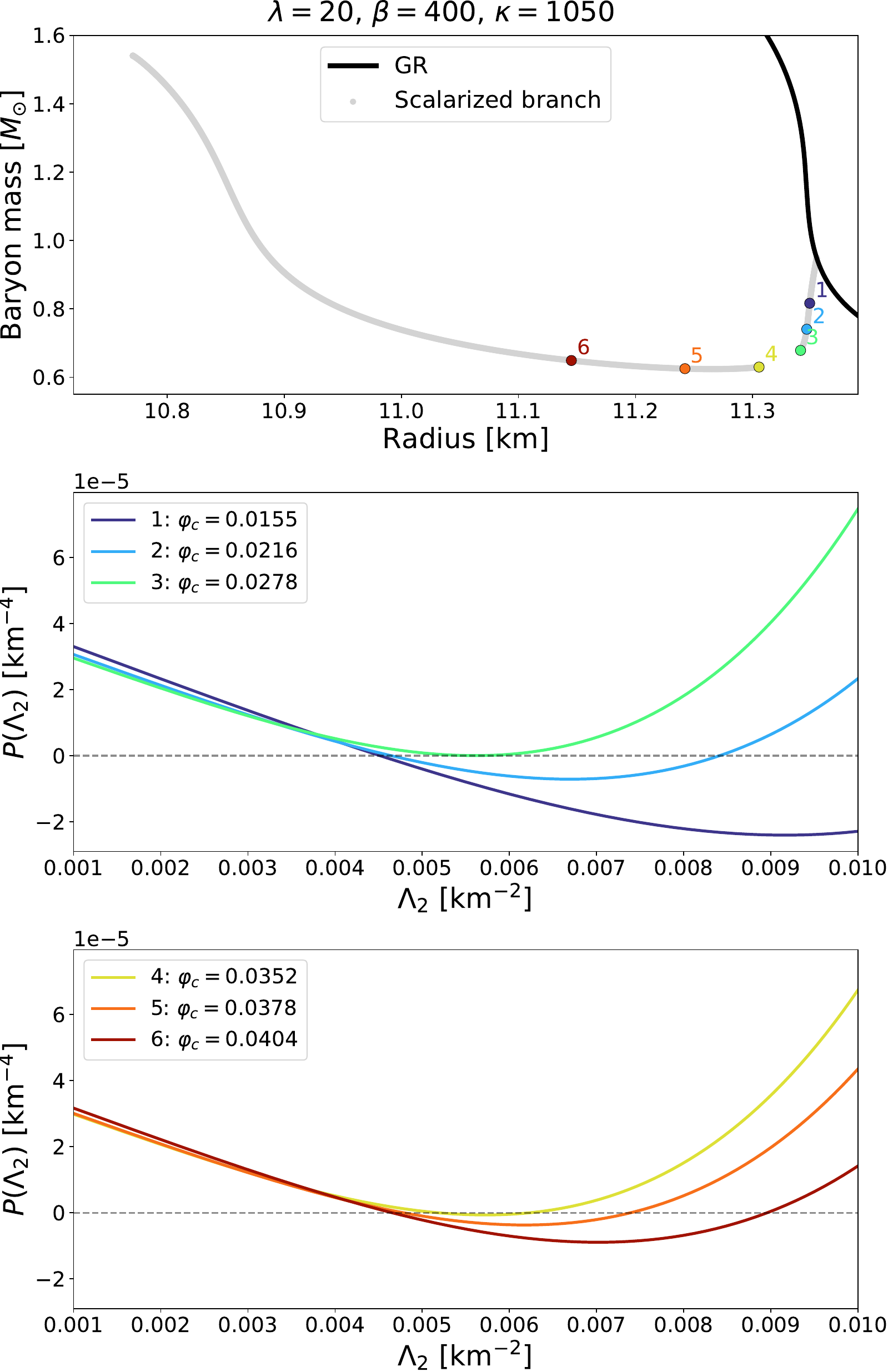}
     \caption{A gapped scalarized branch where terminal points 3 is successfully explained by the violation of the central existence condition. As we build the polynomial for $\Lambda_2$ in Eq.~\eqref{eq: central condition} for scalarized solutions successively closer to points 3 we see that its roots approach each other, leaving us with no roots in the gap (middle panel). The situation looks similar for the other termination point, 4, but the central existence condition is not as close to violation in this case. It will later be revealed in Fig.~\ref{fig: both existence} that the non-central existence condition causes the termination.}
    \label{fig: central condition almost explaining gap}
\end{figure}
The two existence conditions confirmed that the gaps and branch terminations in Sec.~\ref{sec: kappa 0} and~\ref{sec: kappa neq 0} are indeed physical. To start with, recall that the main suspected culprit for the termination of branches was the central existence condition~\eqref{eq: central condition} until this study. Fig.~\ref{fig: central condition almost explaining gap} shows a case where this suspicion is confirmed, but partially. That is, the termination point 3 in Fig.~\ref{fig: central condition almost explaining gap}, the one we knew before the discovery of the reappearance of the branch, suffers from the violation of the central existence condition. Namely, the polynomial $P(\Lambda_2)$ in Eq.~\ref{eq: central condition} has roots which approach each other and then disappear (within numerical precision) as we approach point 3, as seen in the middle panel of Fig.~\ref{fig: central condition almost explaining gap}. However, the point 4 where the branch reappears has a different story. The bottom panel shows that the central existence condition is close to being violated, but it is not exactly violated.

\begin{figure}[t]
    \centering
    \includegraphics[width=0.5\textwidth]{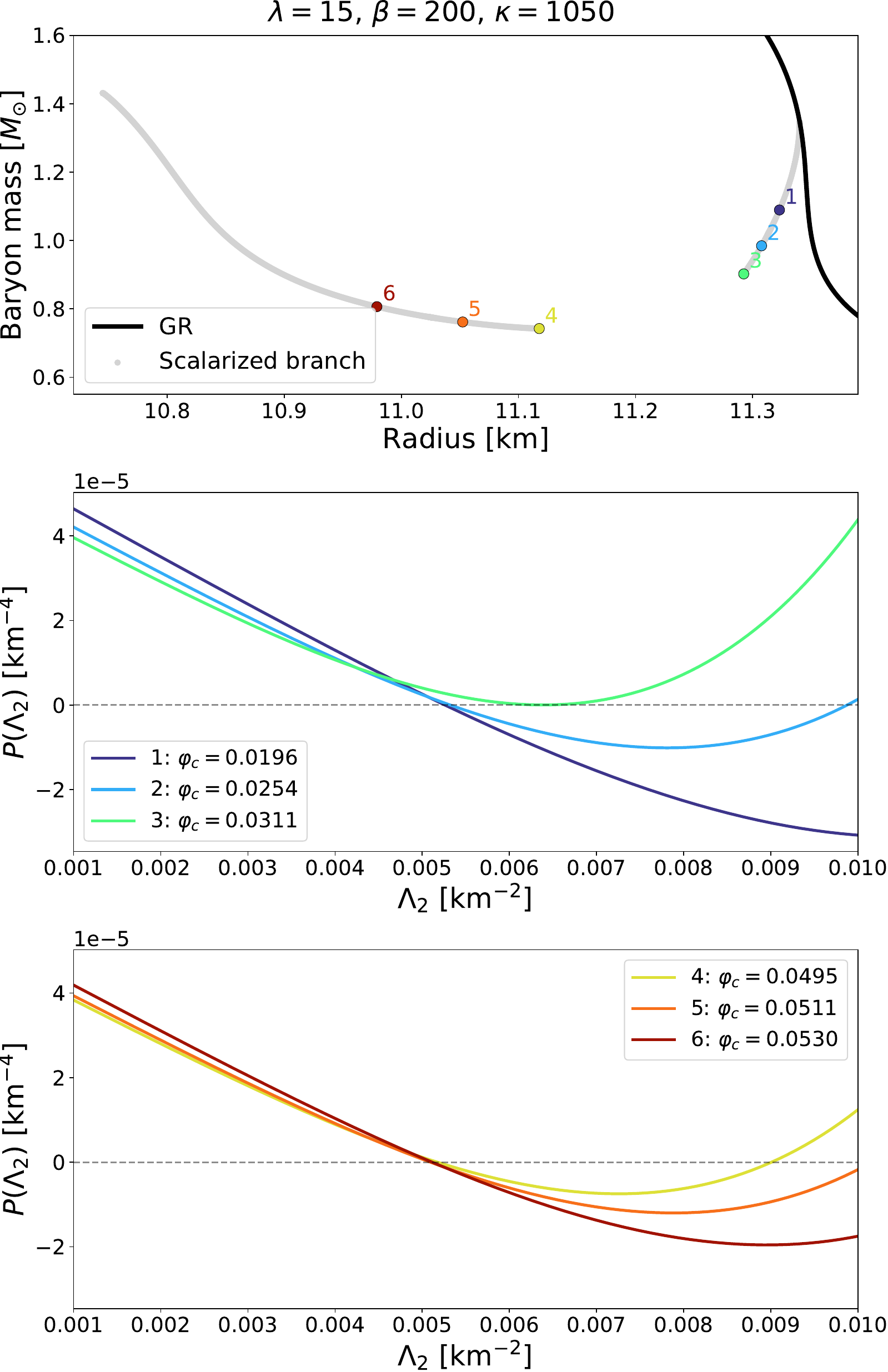}
     \caption{A gapped scalarized branch where one end of the gap, point 4, is clearly not caused by the violation of the central existence condition. Namely, the polynomial $P(\Lambda_2)$ is not losing its roots as we approach this point (bottom panel), meaning either there is a solution we are missing or there is another issue that prevents the existence of the solutions. The latter is the case in the form of the non-central existence condition of Eq.~\eqref{eq: noncentral existence condition}. Termination at point 3 is successfully explained by the central existence condition as before (middle panel).}
    \label{fig: centrail condition NOT explaining gap}
\end{figure}
We can better observe the partial explanatory power of the central existence condition in Fig.~\ref{fig: centrail condition NOT explaining gap} which is a case with different theory parameters. Again, the termination point 3 can be understood in terms of the central existence condition (the middle panel), but not the branch reappearance point 4, where the polynomial is quite far away from not having roots in the bottom panel.

\begin{figure}
    \centering
    \includegraphics[width=\columnwidth]{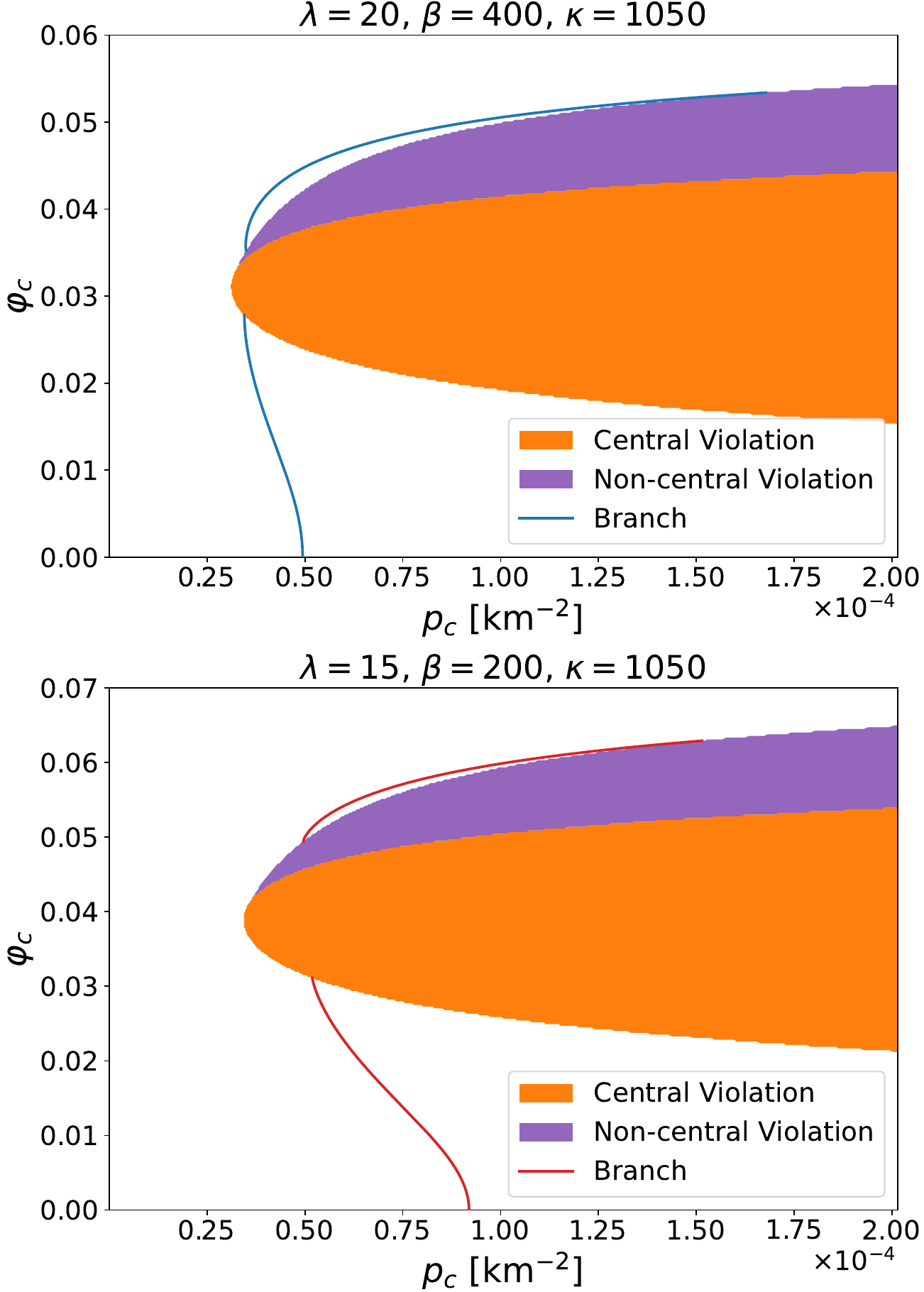}
     \caption{The existence conditions and the scalarized branch for the two different sets of theory parameter values in Figs.~\ref{fig: central condition almost explaining gap} (upper panel) and~\ref{fig: centrail condition NOT explaining gap} (lower panel). The shaded regions are where the existence conditions are violated as marked in the legend. In both cases, the solutions cease to exist, i.e. the branches terminate, once they  enter a region where one of the existence conditions are violated, the termination point at the lower (upper) $\varphi_{\rm c}$ being due to the central (non-central) existence condition.
     } 
    \label{fig: both existence}
\end{figure}
The role of the non-central existence condition can be more explicitly seen in Fig.~\ref{fig: both existence}, which show the regions where either of the existence conditions in Sec.~\ref{sec: nonexistence methods} are violated. Here, we scan the whole $(\rho_0,\varphi_0)$ parameter plane via brute-force, and attempt to construct a star by solving the modified TOV equations at every point. Sometimes, this attempt would fail right at the beginning if the central existence condition is violated. Sometimes, the TOV integration can start successfully, but the differential equation system will become singular at a later point inside the star, which is the violation of the non-central existence condition. Then, we shade these violation regions, and compare it to the termination points of the scalarized branch which is also plotted on the same figure.

Fig.~\ref{fig: both existence} clearly depicts that both termination points correspond to the boundary of the regions where one of the existence conditions are violated. We should emphasize that this is quite a nontrivial finding despite the simplicity of our explanation. Before our discovery of the non-central existence condition, we could only understand one of the terminal points in terms of an existence condition, hence whether we had a purely numerical failure at the other end was still an open question. Discovery of the non-central existence condition is what ensures the physical reality of the gaps and possible physical phenomena that accompany them. To demonstrate the non-triviality of the issue, we  also note that the existence conditions are not an integral part of scalarization in general. The widely studied DEF model, for example, has no analog for them.

We do not present our analysis for all individual cases, but the branch termination points in all our figures coincided with violation of one of the existence conditions, similarly to Fig.~\ref{fig: both existence}. The non-central existence condition is commonly violated at higher values of central scalar field amplitudes $\varphi_{\rm c}$ as in Fig.~\ref{fig: both existence}, hence it is the relevant condition in most cases where scalarized solutions reappear to form a gapped branch.

\subsection{Solutions for the oppositely signed coupling function}
\label{sec: opposite sign}
\begin{figure}
    \centering
    \includegraphics[width=\columnwidth]{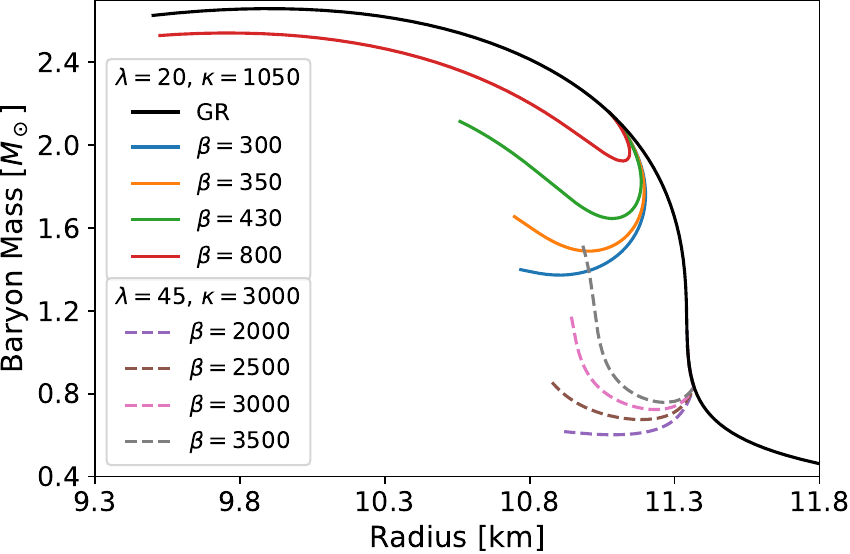}
     \caption{Sample scalarized branches for the oppositely signed $f_{+}$ coupling in Eq.~\eqref{eq:coupling_f1} (for $\epsilon=+1$). Even though we observe early termination of branches, we could not identify any case where there is a gap on a solution branch.}
    \label{fig:f2}
\end{figure}
We noted in Sec.~\ref{sec: sGB gravity} that ${\cal R}^2_{GB}$ can have either sign for neutron stars, hence the tachyonic instability and the resulting scalarization can occur for both signs of the coupling function in~\eqref{eq:coupling_f1}, namely $f_\pm$ for $\epsilon=\pm 1$. In this subsection we will concentrate on $\epsilon=+1$.
This choice of coupling is more prominently studied for black hole scalarization in sGB since non-rotating black holes only scalarize for $\epsilon=+1$~\cite{Doneva:2017bvd,Silva:2017uqg,Doneva:2022ewd}. However, neutron star scalarization is known to have some peculiar features in this case. 

\begin{figure}
    \centering
    \includegraphics[width=\columnwidth]{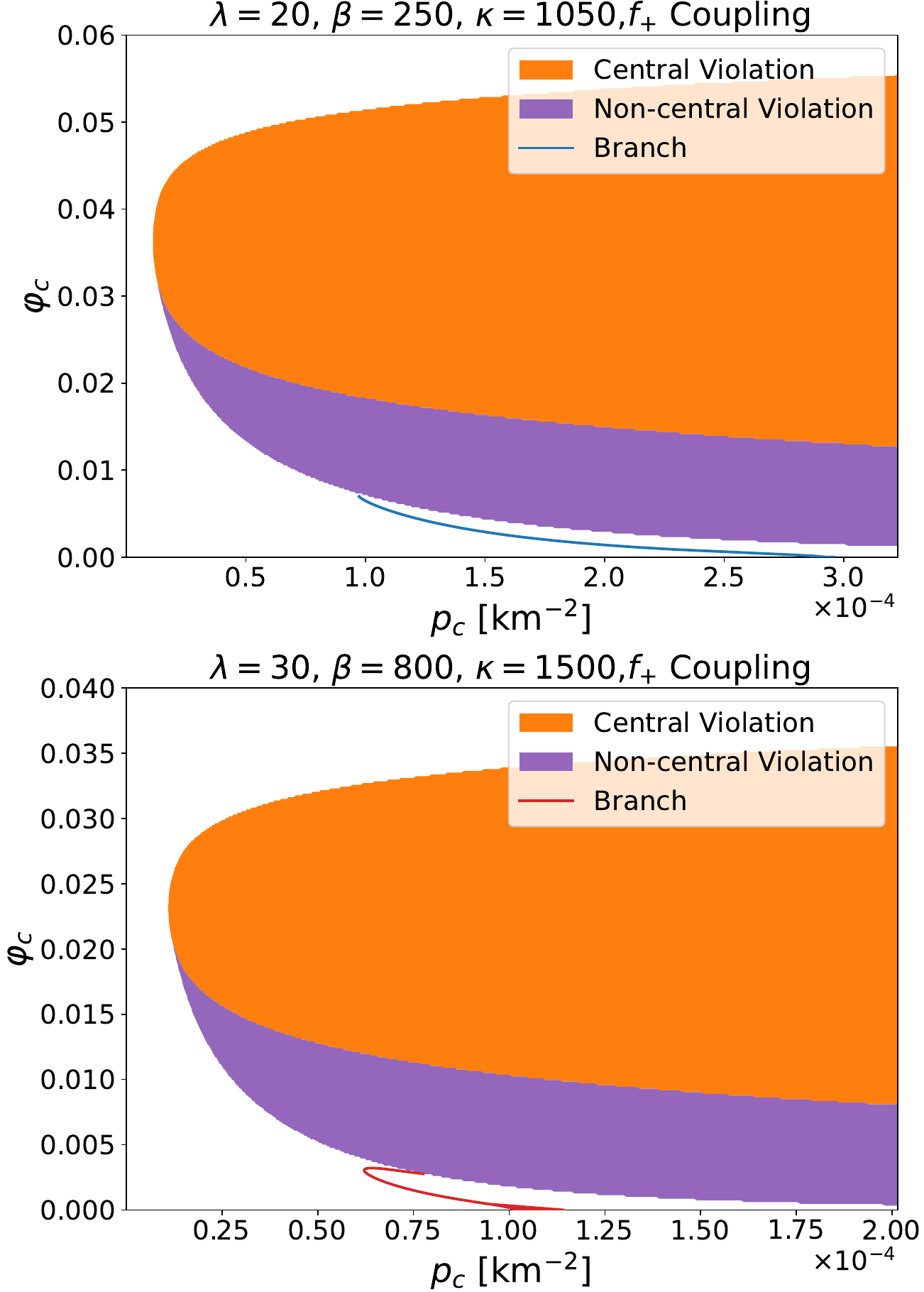}
     \caption{Analog of Fig.~\ref{fig: both existence} for the $f_{+}$ coupling in Eq.~\eqref{eq:coupling_f1}, i.e. with $\epsilon=+1$. The shaded forbidden regions due to the central and non-central existence conditions changed their order, the former occurring at higher $\varphi_{\rm c}$ values than the latter, which is the reverse of $f_-$ (with $\epsilon=-1$) in Fig.~\ref{fig: both existence}. Branches never reappeared at higher $\varphi_{\rm c}$ in the parameter range we searched, hence there were no gapped branches.}
    \label{fig:f2 determinant}
\end{figure}
For $f_+$, the amplitude of the scalar field at the center of the star is typically much smaller compared to scalarization under $f_-$, but the field profile can grow to reach relatively high values off-center~\cite{Doneva:2023kkz}. Recall that one motivation to expect the reappearance of a terminated branch in Sec.~\ref{sec: kappa 0} was based on the expectation that scalarized solutions with higher central scalar field amplitudes $\varphi_{\rm c}$ could exist. Hence, the strong suppression of $\varphi_{\rm c}$ for $f_+$ makes this mechanism less likely to work.

This line of reasoning is indeed realized in the solution branches. We found early terminated branches for certain theory parameter regions of $f_+$, but our searches did not reveal branch reappearance in any case. In other words, we could not find gapped branches for any theory parameter value for $f_+$. Sample scalarized branches can be seen in Fig.~\ref{fig:f2}.

The role of existence conditions are also different for $f_+$ as can be seen in Fig.~\ref{fig:f2 determinant}. The noncentral existence condition becomes relevant at high values of $\varphi_{\rm c}$ for $f_-$ as can be seen in Fig.~\ref{fig: both existence}, however, this is reversed for $f_+$ in Fig.~\ref{fig:f2 determinant}. This is likely related to the fact that the scalar field does not attain its highest amplitude at the center, and grows with radius up to a point for $f_+$, hence violating the non-central existence condition becomes easier.

\section{Discussion}
\label{sec: discussion}

This work examined the spontaneous scalarization of neutron stars in sGB gravity with a specific emphsis on the phase transition picture. During our study we found two novel aspects of the scalarized phase. First, we demonstrated the existence of what we call gaps in the scalarized solution branches, which form a distinct feature of the sGB theory compared to other forms of scalarization investigated so far. Second, in addition to the already known central existence condition, we discovered a novel regularity condition that causes branch termination: the non-central existence condition.  

The power of the Landau theory in understanding generic phase transitions is validated in our case since scalarization in sGB gravity can be understood in similar terms to the DEF model which has been previously analyzed in detail~\cite{Unluturk:2025zie,Ozinan:2026bne}. Many cases in sGB gravity can be interpreted as first- and second-order scalarization in a relatively straightforward manner. However, the more interesting cases are the ones where there are deviations from the standard picture, and the phase transition interpretation was the main driving force in our discovery of such features. For example, early branch termination has already been known in the literature, including the extreme cases where the branch terminates soon after the bifurcation point, as in the low $\beta$ cases in Fig.~\ref{fig: M vs radius kappa=0}. The phase transition point of view suggests to interpret these as a failure to realize a first-order transition. This, in turn, brought the question of whether the branch might later reappear on the mass-radius plot rather than terminating permanently. This led us to the discovery of the gapped branches and their dependence on the theory parameters, somewhat in a similar manner to previous uses of phase transition concepts in revealing scalarized solutions that are easily missed~\cite{Staykov:2026ojk}. 

Any claim about the non-existence of solutions as in branch termination requires stringent standards to ensure that the results are physical and not due to a shortcoming of the computation methods. This led us to the discovery of the non-central existence condition when we realized that the reappearance of a scalarized branch cannot always be explained in terms of the central existence condition. There is an intimate connection between the two conditions as we explain in the Appendix. Furthermore, an understanding of their dependence on the coupling function provides a general idea about which parts of the theory parameter space can be expected to feature reappearance of terminated branches. This also enables us to interpret why there are no gapped branches when we use $f_+$, the coupling function with $\varepsilon=+1$ in Eq.~\eqref{eq:coupling_f1}.

Beyond qualitative predictions, we also investigated how theory parameters determine branch termination and gap formation. The trends can be understood in general terms via the behavior of the existence conditions, which can also provide more universal lessons that can be used for generic coupling functions.

It would be interesting to see how the the existence conditions we studied relate to the well-posedness problems of the sGB theory. It is now well-known that sGB gravity can suffer from loss of hyperbolicity for certain coupling parameters, meaning its time evolution becomes ill-defined after a finite duration both at the linearized~\cite{Blazquez-Salcedo:2020rhf,Blazquez-Salcedo:2020caw} and nonlinear~\cite{East:2021bqk,R:2022hlf,Doneva:2023oww} levels. In the present paper, we were only concerned with stationary solutions, hence such problems do not show up in our results. A detailed examination of which branches can posses a well-posed formulation and how this relates to the termination points is an important follow-up study.


\acknowledgments
K.S and S.Y. acknowledge the partial support of KP-06-N62/6 from the Bulgarian science fund. F.M.R is supported by the Scientific and Technological Research Council of Turkey (T\"UB\.ITAK) Grant Number 122F097. D.D. acknowledges financial support via an Emmy Noether Research Group funded by the German Research Foundation (DFG) under grant no. DO 1771/1-1, by the Spanish Ministry of Science and Innovation via the Ram\'on y Cajal programme (grant RYC2023-042559-I), funded by MCIN/AEI/ 10.13039/501100011033, and by the Spanish Agencia Estatal de Investigaci\'on (grant PID2024-159689NB-C21) funded by the Ministerio de Ciencia, Innovaci\'on y Universidades. We acknowledge Discoverer PetaSC and EuroHPC JU for awarding this project access to Discoverer supercomputer resources. Calculations were also carried out on HEMUS supercomputer at the Center of Excellence in Informatics and ICT. Part of the numerical calculations were performed at KUACC facilities of Koç University. The work was supported by the CoE of Informatics and ICT under Grant No. BG16RFPR002-1.014-0018-C01, financed by PRIDST 2021–2027 and co-financed by the European Union.

\appendix
\renewcommand{\theequation}{A.\arabic{equation}}
\setcounter{equation}{0}
\numberwithin{equation}{section}

\section{An alternative form of the central existence condition}\label{app: central_existence other form}
Here, we show that our central existence condition Eq.~\eqref{eq: central condition} leads to the regularity condition in Eq.~(31) of \textcite{Pani:2011xm}:
\begin{multline}
\alpha^2\beta^2
\leq\frac{1}{7776\pi p_c^4\rho_c}\\
\times\Bigl[128\rho_c^3-27p_c^2\rho_c+288p_c\rho_c^2+54p_c^3\\
-2\sqrt{(3p_c+\rho_c)(3p_c-8\rho_c)^2(3p_c+4\rho_c)^3}\Bigr],
\label{eq:existence_condition_pani}
\end{multline}
This was constructed for a dilatonic sGB theory that has the coupling function $f_{\rm d}(\bar{\varphi})= \frac{\alpha}{16\pi} e^{\beta \bar{\varphi}}$, where $\bar{\varphi}$ is a scaled version of our scalar field $\varphi$.\footnote{\textcite{Pani:2011xm} denote their scalar with $\Phi$, but we avoid this symbol since we already use it for another variable.}

Let us first define
\begin{equation}
C\equiv \bar{\lambda}^4\left[\frac{df}{d\varphi}(\varphi_c)\right]^2\geq 0,
\label{eq:C_def}
\end{equation}
after which, our central existence condition~\eqref{eq: central condition} reads
\begin{equation}
    P(\Lambda_2)\equiv{}9C\Lambda_2^4+72\pi p_cC\Lambda_2^3 -6\pi\rho_c\Lambda_2+16\pi^2\rho_c^2=0,
\label{eq:center_quartic}
\end{equation}
We are only interested in the region $\Lambda_2 >0$, where we have the following properties.
\begin{itemize}
    \item Since $P''(\Lambda_2)>0$, $P(\Lambda_2)$ is strictly convex, 
    \item Since $P'(0)<0$, $P(\Lambda_2)$ has a single minimum.
    \item Since $P(0)>0$, Eq.~\eqref{eq:center_quartic} has two, one (double), or no positive roots, depending on whether this minimum is negative, zero, or positive.
\end{itemize}
This means the number of the roots will change with $C$. 

Let's call the critical value where the two roots merge $C_{\rm crit}$. Since $\partial P/\partial C=9\Lambda_2^3(\Lambda_2+8\pi p_c)>0$, a positive root exists if and only if $C\leq C_{\rm crit}$.  In other words, a static, scalarized neutron star solution can exist only if
\begin{equation}
    \bar{\lambda}^4\left[\frac{df}{d\varphi}(\varphi_c)\right]^2\leq C_{\rm crit} .
\label{eq:general_existence_condition}
\end{equation}
This is essentially Eq.~(31) of \textcite{Pani:2011xm}. The only remaining part is finding $C_{\rm crit}$.

At $C = C_{\rm crit}$, there is a double root $\Lambda_{2,\rm crit}$, which satisfies
\begin{equation}
P(\Lambda_{2,\rm crit})=0=P'(\Lambda_{2,\rm crit}).
\label{eq:double_root}
\end{equation}
$0=P'(\Lambda_{2,\rm crit})$ gives
\begin{equation}
C_{\rm crit}=\frac{\pi\rho_c}
{6\Lambda_{2,\rm crit}^2\left(\Lambda_{2,\rm crit}+6\pi p_c\right)}.
\label{eq:Ccrit_intermediate}
\end{equation}
Substituting this into $P(\Lambda_{2,\rm crit})=0$ leads to
\begin{equation}
    9\Lambda_{2,\rm crit}^2+16\pi(3p_c-2\rho_c)\Lambda_{2,\rm crit}-192\pi^2p_c\rho_c=0.
\label{eq:Lambda2crit_quadratic}
\end{equation}
The only positive root of this equation is
\begin{align}
    \Lambda_{2,\rm crit}&=\frac{8\pi}{9}\left(2\rho_c-3p_c+s\right),
\label{eq:Lambda2crit}
\end{align}
where $s\equiv\sqrt{(3p_c+\rho_c)(3p_c+4\rho_c)}$. Inserting this into Eq.~\eqref{eq:Ccrit_intermediate}, we finally find the critical value as
\begin{equation}
    C_{\rm crit}={}\frac{243\rho_c}{256\pi^2\left(2\rho_c-3p_c+s\right)^2 \left(15p_c+8\rho_c+4s\right)} .
\label{eq:Ccrit}
\end{equation}

The only remaining part is inserting $C_{\rm crit}$ above into Eq.~\eqref{eq:general_existence_condition} and organize the terms:
\begin{multline}
    8\pi\bar{\lambda}^4\left[\frac{df}{d\varphi}(\varphi_c)\right]^2
    \leq\frac{1}{7776\pi p_c^4\rho_c}\\
    \times\Bigl[128\rho_c^3-27p_c^2\rho_c+288p_c\rho_c^2+54p_c^3\\
    -2(8\rho_c-3p_c)\sqrt{(3p_c+\rho_c)(3p_c+4\rho_c)^3}\Bigr] .
\label{eq:existence_condition_general_form}
\end{multline}
However, this is still not in the exact same form of Eq.~\eqref{eq:existence_condition_pani}. The main difference is due to the fact that Eq.~\eqref{eq:existence_condition_general_form} holds for all $p_c,\rho_c>0$. If $3p_c\leq8\rho_c$, which is satisfied by all matter obeying the dominant energy condition $p_c\leq\rho_c$, then $8\rho_c-3p_c=\sqrt{(3p_c-8\rho_c)^2}$. When the coupling function is given by $f_{\rm d}(\bar{\varphi})= \frac{\alpha}{16\pi} e^{\beta \bar{\varphi}}$, $\bar{\varphi}$ is scaled properly, and we use the assumption $\bar{\varphi} \ll 1 $ in \textcite{Pani:2011xm}, Eq.~\eqref{eq:existence_condition_general_form} finally reduces to Eq.~\eqref{eq:existence_condition_pani}.

\section{Relationship between the existence conditions}\label{app: two conditions are related}
In this appendix, we demonstrate that the central and non-central existence conditions presented in Sec.~\ref{sec: nonexistence methods} are closely related. The algebraic derivations in this section were performed with the assistance of ChatGPT version 5.6 by OpenAI, and verified by the authors.

For convenience, let us introduce the notation
\begin{equation}
    E \equiv e^{-2\Lambda},
    \qquad
    F(\varphi) \equiv \bar{\lambda}^2 \frac{df}{d\varphi},
    \qquad
    X \equiv \frac{d\varphi}{dr}.
\end{equation}

A tedious but simple calculation gives the determinant of our radial evolution system in Eq.~\eqref{eq:M_radial_evolution}, $\mathcal{M}(r)\ U' = S(U)$, as
\begin{equation}
    \det{\cal M}
    =
    \frac{2}{r^5}\,{\cal D},
\end{equation}
where
\begin{align}
{\cal D}
={}&
\left(r^2-4EFrX\right)^2
-4F^2(1-E)^2
\nonumber\\
&+
16EF^2(1-E)
\left[
    (1+3E)FX-r
\right]\Phi'.
\label{eq:D_definition}
\end{align}
Thus,
\begin{equation}
    {\cal D}(r)\neq0
\label{eq:D_condition}
\end{equation}
is an additional local non-degeneracy condition for $r \neq 0$, what we called the non-central existence condition.

To see the relationship of this to the central existence condition, let us first recall that we can rewrite the central existence condition~\eqref{eq: central condition} as
\begin{equation}
P(\Lambda_2)
=
9F_0^2\Lambda_2^4
+
72\pi p_0F_0^2\Lambda_2^3
-
6\pi\rho_0\Lambda_2
+
16\pi^2\rho_0^2  \,,
\label{eq:center_quartic2}
\end{equation}
where $F_0 \equiv \bar{\lambda}^2 \left. \sfrac{df}{d\varphi} \right|_{\varphi=\varphi_0}$ is the first term in our usual series expansion around $r=0$. Also note that $F_0^2 = C$ in Eq.~\eqref{eq:C_def}. Even though we emphasized the vanishing of this polynomial as the condition, a closer look shows that the central existence condition is better expressed as 
\begin{equation}
    P(\Lambda_2)=0,
    \qquad
    \frac{dP}{d\Lambda_2}(\Lambda_2) = P'(\Lambda_2) \neq 0.
\end{equation}
The second equation avoids the merger of the two roots into one, algebraic degeneracy. This point was discussed in Appendix~\ref{app: central_existence other form}. Moreover, we explicitly showed in the main text that this merger is indeed the case at the point where the branches terminate due to the central existence condition in Sec.~\ref{sec: nonexistence methods}.

A relationship between this algebraic degeneracy and the determinant of the radial integration equations appears when we series-expand the determinant and rearrange the terms to arrive at
\begin{equation}
    {\cal D}(r) = \overbrace{\left[ \frac{64\pi^2\rho_0^2}{9\Lambda_2^2}
    -
    12F_0^2\Lambda_2^2
    -
    64\pi p_0F_0^2\Lambda_2 \right]}^{{\cal D}_4} r^4 + \mathcal{O}(r^6) .
\end{equation}
The connection becomes clear once we recongnize that
\begin{equation}
    {\cal D}_4
    =
    -\frac{4}{9\Lambda_2}
    P'(\Lambda_2).
\label{eq:D0_Pprime}
\end{equation}
Hence, encountering a double root at the central existence condition and having a vanishing determinant at the non-central existence condition are both related to one single equality: $P'(\Lambda_2)=0$.

We should be clear that the two conditions are related, but they are not identical. The above relationship is based on what happens to ${\cal D}$ at $r=0$, but we use ${\cal D}$ specifically at $r \neq 0$ when considering the non-central existence condition. Nevertheless, we can think of ${\cal D}_4=0$ to be the case where the violation of the non-central existence condition appears for the first time. This violation then moves to higher radii as the initial value $p_c$ in the integration changes, to provide a novel constraint on scalarization.

Finally, we should also recall that these results are derived under the assumption of perfect analytical integration of the differential equation system. One needs to consider the conditioning of the matrices and the fact that we cannot observe ${\cal D}(r)=0$ exactly in numerical computations, as discussed in Sec.~\ref{sec: nonexistence methods}.

\bibliography{references}

@article{Kuan:2022oxs,
    author = "Kuan, Hao-Jui and Suvorov, Arthur G. and Doneva, Daniela D. and Yazadjiev, Stoytcho S.",
    title = "{Gravitational Waves from Accretion-Induced Descalarization in Massive Scalar-Tensor Theory}",
    eprint = "2203.03672",
    archivePrefix = "arXiv",
    primaryClass = "gr-qc",
    reportNumber = "ET-0034A-22",
    doi = "10.1103/PhysRevLett.129.121104",
    journal = "Phys. Rev. Lett.",
    volume = "129",
    number = "12",
    pages = "121104",
    year = "2022"
}

@article{Glendenning:1998ag,
    author = "Glendenning, Norman K. and Kettner, Christiane",
    title = "{Nonidentical neutron star twins}",
    eprint = "astro-ph/9807155",
    archivePrefix = "arXiv",
    reportNumber = "LBL-42080, LBNL-42080",
    journal = "Astron. Astrophys.",
    volume = "353",
    pages = "L9",
    year = "2000"
}

@article{Zhao:2022vig,
    author = "Zhao, Junjie and Freire, Paulo C. C. and Kramer, Michael and Shao, Lijing and Wex, Norbert",
    title = "{Closing a spontaneous-scalarization window with binary pulsars}",
    eprint = "2201.03771",
    archivePrefix = "arXiv",
    primaryClass = "astro-ph.HE",
    doi = "10.1088/1361-6382/ac69a3",
    journal = "Class. Quant. Grav.",
    volume = "39",
    number = "11",
    pages = "11LT01",
    year = "2022"
}

@article{LIGOScientific:2018dkp,
    author = "Abbott, B. P. and others",
    collaboration = "LIGO Scientific, Virgo",
    title = "{Tests of General Relativity with GW170817}",
    eprint = "1811.00364",
    archivePrefix = "arXiv",
    primaryClass = "gr-qc",
    reportNumber = "LIGO-P1800059",
    doi = "10.1103/PhysRevLett.123.011102",
    journal = "Phys. Rev. Lett.",
    volume = "123",
    number = "1",
    pages = "011102",
    year = "2019"
}

@article{Silva:2020acr,
    author = "Silva, Hector O. and Holgado, A. Miguel and C\'ardenas-Avenda\~no, Alejandro and Yunes, Nicol\'as",
    title = "{Astrophysical and theoretical physics implications from multimessenger neutron star observations}",
    eprint = "2004.01253",
    archivePrefix = "arXiv",
    primaryClass = "gr-qc",
    doi = "10.1103/PhysRevLett.126.181101",
    journal = "Phys. Rev. Lett.",
    volume = "126",
    number = "18",
    pages = "181101",
    year = "2021"
}

@article{Charmousis:2021npl,
    author = "Charmousis, Christos and Leh{\'e}bel, Antoine and Smyrniotis, Evangelos and Stergioulas, Nikolaos",
    title = "{Astrophysical constraints on compact objects in 4D Einstein-Gauss-Bonnet gravity}",
    eprint = "2109.01149",
    archivePrefix = "arXiv",
    primaryClass = "gr-qc",
    doi = "10.1088/1475-7516/2022/02/033",
    journal = "JCAP",
    volume = "02",
    number = "02",
    pages = "033",
    year = "2022"
}

@article{Soldateschi:2020hju,
    author = "Soldateschi, J. and Bucciantini, N. and Del Zanna, L.",
    title = "{Axisymmetric equilibrium models for magnetised neutron stars in Scalar-Tensor Theories}",
    eprint = "2005.12758",
    archivePrefix = "arXiv",
    primaryClass = "astro-ph.HE",
    doi = "10.1051/0004-6361/202037918",
    journal = "Astron. Astrophys.",
    volume = "640",
    pages = "A44",
    year = "2020"
}

@article{Yazadjiev:2015zia,
    author = "Yazadjiev, Stoytcho S. and Doneva, Daniela D. and Kokkotas, Kostas D.",
    title = "{Rapidly rotating neutron stars in R-squared gravity}",
    eprint = "1501.04591",
    archivePrefix = "arXiv",
    primaryClass = "gr-qc",
    doi = "10.1103/PhysRevD.91.084018",
    journal = "Phys. Rev. D",
    volume = "91",
    number = "8",
    pages = "084018",
    year = "2015"
}

@article{Kleihaus:2016dui,
    author = "Kleihaus, Burkhard and Kunz, Jutta and Mojica, Sindy and Zagermann, Marco",
    title = "{Rapidly Rotating Neutron Stars in Dilatonic Einstein-Gauss-Bonnet Theory}",
    eprint = "1601.05583",
    archivePrefix = "arXiv",
    primaryClass = "gr-qc",
    doi = "10.1103/PhysRevD.93.064077",
    journal = "Phys. Rev. D",
    volume = "93",
    number = "6",
    pages = "064077",
    year = "2016"
}

@article{Stefanov:2007eq,
    author = "Stefanov, Ivan Zh. and Yazadjiev, Stoytcho S. and Todorov, Michail D.",
    title = "{Phases of 4D scalar-tensor black holes coupled to Born-Infeld nonlinear electrodynamics}",
    eprint = "0708.4141",
    archivePrefix = "arXiv",
    primaryClass = "gr-qc",
    doi = "10.1142/S0217732308028351",
    journal = "Mod. Phys. Lett. A",
    volume = "23",
    pages = "2915--2931",
    year = "2008"
}

@article{Doneva:2013qva,
    author = "Doneva, Daniela D. and Yazadjiev, Stoytcho S. and Stergioulas, Nikolaos and Kokkotas, Kostas D.",
    title = "{Rapidly rotating neutron stars in scalar-tensor theories of gravity}",
    eprint = "1309.0605",
    archivePrefix = "arXiv",
    primaryClass = "gr-qc",
    doi = "10.1103/PhysRevD.88.084060",
    journal = "Phys. Rev. D",
    volume = "88",
    number = "8",
    pages = "084060",
    year = "2013"
}

@article{Horbatsch:2015bua,
    author = "Horbatsch, Michael and Silva, Hector O. and Gerosa, Davide and Pani, Paolo and Berti, Emanuele and Gualtieri, Leonardo and Sperhake, Ulrich",
    title = "{Tensor-multi-scalar theories: relativistic stars and 3 + 1 decomposition}",
    eprint = "1505.07462",
    archivePrefix = "arXiv",
    primaryClass = "gr-qc",
    doi = "10.1088/0264-9381/32/20/204001",
    journal = "Class. Quant. Grav.",
    volume = "32",
    number = "20",
    pages = "204001",
    year = "2015"
}

@article{Doneva:2019ltb,
    author = "Doneva, Daniela D. and Yazadjiev, Stoytcho S.",
    title = "{Topological neutron stars in tensor-multi-scalar theories of gravity}",
    eprint = "1911.06908",
    archivePrefix = "arXiv",
    primaryClass = "gr-qc",
    doi = "10.1103/PhysRevD.101.064072",
    journal = "Phys. Rev. D",
    volume = "101",
    number = "6",
    pages = "064072",
    year = "2020"
}

@article{Blazquez-Salcedo:2020caw,
    author = "Bl{\'a}zquez-Salcedo, Jose Luis and Doneva, Daniela D. and Kahlen, Sarah and Kunz, Jutta and Nedkova, Petya and Yazadjiev, Stoytcho S.",
    title = "{Polar quasinormal modes of the scalarized Einstein-Gauss-Bonnet black holes}",
    eprint = "2006.06006",
    archivePrefix = "arXiv",
    primaryClass = "gr-qc",
    doi = "10.1103/PhysRevD.102.024086",
    journal = "Phys. Rev. D",
    volume = "102",
    number = "2",
    pages = "024086",
    year = "2020"
}

@article{Blazquez-Salcedo:2020rhf,
    author = "Bl{\'a}zquez-Salcedo, Jose Luis and Doneva, Daniela D. and Kahlen, Sarah and Kunz, Jutta and Nedkova, Petya and Yazadjiev, Stoytcho S.",
    title = "{Axial perturbations of the scalarized Einstein-Gauss-Bonnet black holes}",
    eprint = "2003.02862",
    archivePrefix = "arXiv",
    primaryClass = "gr-qc",
    doi = "10.1103/PhysRevD.101.104006",
    journal = "Phys. Rev. D",
    volume = "101",
    number = "10",
    pages = "104006",
    year = "2020"
}

@article{Yagi:2013mbt,
    author = "Yagi, Kent and Stein, Leo C. and Yunes, Nicolas and Tanaka, Takahiro",
    title = "{Isolated and Binary Neutron Stars in Dynamical Chern-Simons Gravity}",
    eprint = "1302.1918",
    archivePrefix = "arXiv",
    primaryClass = "gr-qc",
    doi = "10.1103/PhysRevD.87.084058",
    journal = "Phys. Rev. D",
    volume = "87",
    pages = "084058",
    year = "2013",
    note = "[Erratum: Phys.Rev.D 93, 089909 (2016)]"
}

@article{Yazadjiev:2016pcb,
    author = "Yazadjiev, Stoytcho S. and Doneva, Daniela D. and Popchev, Dimitar",
    title = "{Slowly rotating neutron stars in scalar-tensor theories with a massive scalar field}",
    eprint = "1602.04766",
    archivePrefix = "arXiv",
    primaryClass = "gr-qc",
    doi = "10.1103/PhysRevD.93.084038",
    journal = "Phys. Rev. D",
    volume = "93",
    number = "8",
    pages = "084038",
    year = "2016"
}

@article{Silva:2017uqg,
    author = "Silva, Hector O. and Sakstein, Jeremy and Gualtieri, Leonardo and Sotiriou, Thomas P. and Berti, Emanuele",
    title = "{Spontaneous scalarization of black holes and compact stars from a Gauss-Bonnet coupling}",
    eprint = "1711.02080",
    archivePrefix = "arXiv",
    primaryClass = "gr-qc",
    doi = "10.1103/PhysRevLett.120.131104",
    journal = "Phys. Rev. Lett.",
    volume = "120",
    number = "13",
    pages = "131104",
    year = "2018"
}

@article{Doneva:2017duq,
    author = "Doneva, Daniela D. and Yazadjiev, Stoytcho S.",
    title = "{Neutron star solutions with curvature induced scalarization in the extended Gauss-Bonnet scalar-tensor theories}",
    eprint = "1712.03715",
    archivePrefix = "arXiv",
    primaryClass = "gr-qc",
    doi = "10.1088/1475-7516/2018/04/011",
    journal = "JCAP",
    volume = "04",
    pages = "011",
    year = "2018"
}

@article{Xu:2021kfh,
    author = "Xu, Rui and Gao, Yong and Shao, Lijing",
    title = "{Neutron stars in massive scalar-Gauss-Bonnet gravity: Spherical structure and time-independent perturbations}",
    eprint = "2111.06561",
    archivePrefix = "arXiv",
    primaryClass = "gr-qc",
    doi = "10.1103/PhysRevD.105.024003",
    journal = "Phys. Rev. D",
    volume = "105",
    number = "2",
    pages = "024003",
    year = "2022"
}

@article{Doneva:2022byd,
    author = "Doneva, Daniela D. and Va{\~n}{\'o}-Vi{\~n}uales, Alex and Yazadjiev, Stoytcho S.",
    title = "{Dynamical descalarization with a jump during a black hole merger}",
    eprint = "2204.05333",
    archivePrefix = "arXiv",
    primaryClass = "gr-qc",
    doi = "10.1103/PhysRevD.106.L061502",
    journal = "Phys. Rev. D",
    volume = "106",
    number = "6",
    pages = "L061502",
    year = "2022"
}

@article{Herdeiro:2026sur,
    author = "Herdeiro, Carlos and Huang, Hyat and Kunz, Jutta and Lai, Meng-Yun and Radu, Eugen and Zou, De-Cheng",
    title = "{Phase structure of scalarized black holes in Einstein-scalar-Gauss-Bonnet gravity}",
    eprint = "2603.24164",
    archivePrefix = "arXiv",
    primaryClass = "gr-qc",
    doi = "10.1103/4k1l-jzvz",
    journal = "Phys. Rev. D",
    volume = "113",
    number = "12",
    pages = "124046",
    year = "2026"
}

@article{Doneva:2021tvn,
    author = "Doneva, Daniela D. and Yazadjiev, Stoytcho S.",
    title = "{Beyond the spontaneous scalarization: New fully nonlinear mechanism for the formation of scalarized black holes and its dynamical development}",
    eprint = "2107.01738",
    archivePrefix = "arXiv",
    primaryClass = "gr-qc",
    doi = "10.1103/PhysRevD.105.L041502",
    journal = "Phys. Rev. D",
    volume = "105",
    number = "4",
    pages = "L041502",
    year = "2022"
}

@article{Ventagli:2021ubn,
    author = "Ventagli, Giulia and Antoniou, Georgios and Leh{\'e}bel, Antoine and Sotiriou, Thomas P.",
    title = "{Neutron star scalarization with Gauss-Bonnet and Ricci scalar couplings}",
    eprint = "2111.03644",
    archivePrefix = "arXiv",
    primaryClass = "gr-qc",
    doi = "10.1103/PhysRevD.104.124078",
    journal = "Phys. Rev. D",
    volume = "104",
    number = "12",
    pages = "124078",
    year = "2021"
}

@article{Akmal:1998cf,
    author = "Akmal, A. and Pandharipande, V. R. and Ravenhall, D. G.",
    title = "{The Equation of state of nucleon matter and neutron star structure}",
    eprint = "nucl-th/9804027",
    archivePrefix = "arXiv",
    doi = "10.1103/PhysRevC.58.1804",
    journal = "Phys. Rev. C",
    volume = "58",
    pages = "1804--1828",
    year = "1998"
}

@article{Read:2008iy,
    author = "Read, Jocelyn S. and Lackey, Benjamin D. and Owen, Benjamin J. and Friedman, John L.",
    title = "{Constraints on a phenomenologically parameterized neutron-star equation of state}",
    eprint = "0812.2163",
    archivePrefix = "arXiv",
    primaryClass = "astro-ph",
    doi = "10.1103/PhysRevD.79.124032",
    journal = "Phys. Rev. D",
    volume = "79",
    pages = "124032",
    year = "2009"
}

@article{Damour:1992we,
    author = "Damour, Thibault and Esposito-Farese, Gilles",
    title = "{Tensor multiscalar theories of gravitation}",
    reportNumber = "IHES-P-91-93, CPT-91-PE-2542",
    doi = "10.1088/0264-9381/9/9/015",
    journal = "Class. Quant. Grav.",
    volume = "9",
    pages = "2093--2176",
    year = "1992"
}

@article{Damour:1993hw,
    author = "Damour, Thibault and Esposito-Farese, Gilles",
    title = "{Nonperturbative strong field effects in tensor - scalar theories of gravitation}",
    reportNumber = "IHES-P-93-1, CPT-93-PE-2868",
    doi = "10.1103/PhysRevLett.70.2220",
    journal = "Phys. Rev. Lett.",
    volume = "70",
    pages = "2220--2223",
    year = "1993"
}

@article{Damour:1996ke,
    author = "Damour, Thibault and Esposito-Farese, Gilles",
    title = "{Tensor - scalar gravity and binary pulsar experiments}",
    eprint = "gr-qc/9602056",
    archivePrefix = "arXiv",
    reportNumber = "IHES-P-96-13, IASSNS-HEP-96-14, BRX-TH-391, CPT-95-PE-3279, CPT-95-P.E.3279",
    doi = "10.1103/PhysRevD.54.1474",
    journal = "Phys. Rev. D",
    volume = "54",
    pages = "1474--1491",
    year = "1996"
}

@article{Doneva:2017bvd,
    author = "Doneva, Daniela D. and Yazadjiev, Stoytcho S.",
    title = "{New Gauss-Bonnet Black Holes with Curvature-Induced Scalarization in Extended Scalar-Tensor Theories}",
    eprint = "1711.01187",
    archivePrefix = "arXiv",
    primaryClass = "gr-qc",
    doi = "10.1103/PhysRevLett.120.131103",
    journal = "Phys. Rev. Lett.",
    volume = "120",
    number = "13",
    pages = "131103",
    year = "2018"
}

@article{Yazadjiev:2025ezx,
    author = "Yazadjiev, Stoytcho S. and Doneva, Daniela D.",
    title = "{No-hair theorems in general relativity and scalar{\textendash}tensor theories}",
    eprint = "2505.01038",
    archivePrefix = "arXiv",
    primaryClass = "gr-qc",
    doi = "10.1142/S0218271825300046",
    journal = "Int. J. Mod. Phys. D",
    volume = "34",
    number = "09",
    pages = "2530004",
    year = "2025"
}

@article{Antoniou:2017acq,
    author = "Antoniou, G. and Bakopoulos, A. and Kanti, P.",
    title = "{Evasion of No-Hair Theorems and Novel Black-Hole Solutions in Gauss-Bonnet Theories}",
    eprint = "1711.03390",
    archivePrefix = "arXiv",
    primaryClass = "hep-th",
    doi = "10.1103/PhysRevLett.120.131102",
    journal = "Phys. Rev. Lett.",
    volume = "120",
    number = "13",
    pages = "131102",
    year = "2018"
}

@article{Doneva:2022ewd,
    author = "Doneva, Daniela D. and Ramazano{\u{g}}lu, Fethi M. and Silva, Hector O. and Sotiriou, Thomas P. and Yazadjiev, Stoytcho S.",
    title = "{Spontaneous scalarization}",
    eprint = "2211.01766",
    archivePrefix = "arXiv",
    primaryClass = "gr-qc",
    doi = "10.1103/RevModPhys.96.015004",
    journal = "Rev. Mod. Phys.",
    volume = "96",
    number = "1",
    pages = "015004",
    year = "2024"
}

@article{Pani:2011xm,
    author = "Pani, Paolo and Berti, Emanuele and Cardoso, Vitor and Read, Jocelyn",
    title = "{Compact stars in alternative theories of gravity. Einstein-Dilaton-Gauss-Bonnet gravity}",
    eprint = "1109.0928",
    archivePrefix = "arXiv",
    primaryClass = "gr-qc",
    doi = "10.1103/PhysRevD.84.104035",
    journal = "Phys. Rev. D",
    volume = "84",
    pages = "104035",
    year = "2011"
}

@article{Yazadjiev:2014cza,
    author = "Yazadjiev, Stoytcho S. and Doneva, Daniela D. and Kokkotas, Kostas D. and Staykov, Kalin V.",
    title = "{Non-perturbative and self-consistent models of neutron stars in R-squared gravity}",
    eprint = "1402.4469",
    archivePrefix = "arXiv",
    primaryClass = "gr-qc",
    doi = "10.1088/1475-7516/2014/06/003",
    journal = "JCAP",
    volume = "06",
    pages = "003",
    year = "2014"
}

@article{Harada:1998ge,
    author = "Harada, Tomohiro",
    title = "{Neutron stars in scalar tensor theories of gravity and catastrophe theory}",
    eprint = "gr-qc/9801049",
    archivePrefix = "arXiv",
    reportNumber = "KUNS-1476",
    doi = "10.1103/PhysRevD.57.4802",
    journal = "Phys. Rev. D",
    volume = "57",
    pages = "4802--4811",
    year = "1998"
}

@article{Staykov:2026ojk,
    author = "Staykov, V., Kalin and Ramazano{\u{g}}lu, Fethi M. and Doneva, Daniela D. and Yazadjiev, Stoytcho S.",
    title = "{Phase transition structure of scalarized neutron stars: The effect of rotation and linear coupling}",
    eprint = "2604.21298",
    archivePrefix = "arXiv",
    primaryClass = "gr-qc",
    doi = "10.1103/8ht3-r4nb",
    journal = "Phys. Rev. D",
    volume = "114",
    number = "2",
    pages = "024082",
    year = "2026"
}

@article{Ozinan:2026bne,
    author = {{\"O}zinan, Murat and {\"U}nl{\"u}t{\"u}rk, K{\i}van{\c{c}} {\.I}. and Ramazano{\u{g}}lu, Fethi M.},
    title = "{Underlying mechanisms of phase transitions in scalar-tensor theories}",
    eprint = "2604.24867",
    archivePrefix = "arXiv",
    primaryClass = "gr-qc",
    doi = "10.1103/1brh-qzls",
    journal = "Phys. Rev. D",
    volume = "114",
    number = "2",
    pages = "024054",
    year = "2026"
}

@misc{Ozinan:2026ffv,
    author = {{\"O}zinan, Murat and Yamak, O{\u{g}}uzhan K. and Ramazano{\u{g}}lu, Fethi M.},
    title = "{Novel Observable Signals from First-Order Gravitational Phase Transitions}",
    eprint = "2608.02736",
    archivePrefix = "arXiv",
    primaryClass = "gr-qc",
    month = "8",
    year = "2026"
}

@article{Unluturk:2025zie,
    author = {{\"U}nl{\"u}t{\"u}rk, K{\i}van{\c{c}} {\.I}. and Tuna, Semih and Yamak, O{\u{g}}uzhan K. and Ramazano{\u{g}}lu, Fethi M.},
    title = "{Nature of Phase Transitions and Metastability in Scalar-Tensor Theories}",
    eprint = "2502.01781",
    archivePrefix = "arXiv",
    primaryClass = "gr-qc",
    doi = "10.1103/14xz-d7xf",
    journal = "Phys. Rev. Lett.",
    volume = "135",
    number = "6",
    pages = "061401",
    year = "2025"
}

@article{Doneva:2023kkz,
    author = {Doneva, Daniela D. and Kr{\"u}ger, Christian J. and Staykov, Kalin V. and Yordanov, Petar Y.},
    title = "{Neutron stars in Gauss-Bonnet gravity: Nonlinear scalarization and gravitational phase transitions}",
    eprint = "2306.16988",
    archivePrefix = "arXiv",
    primaryClass = "gr-qc",
    doi = "10.1103/PhysRevD.108.044054",
    journal = "Phys. Rev. D",
    volume = "108",
    number = "4",
    pages = "044054",
    year = "2023"
}

@article{Muniz:2025egq,
    author = "Muniz, Jo{\~a}o Vitor M. and Ortiz, N{\'e}stor and Mendes, Raissa F. P.",
    title = "{Phase transition mechanism of spontaneous scalarization}",
    eprint = "2503.11385",
    archivePrefix = "arXiv",
    primaryClass = "gr-qc",
    doi = "10.1103/qh47-qv76",
    journal = "Phys. Rev. D",
    volume = "112",
    number = "6",
    pages = "064037",
    year = "2025"
}

@misc{M:2026yut,
    author = "M., Jos{\'e} Carlos Olvera and Doneva, Daniela D. and Cerd{\'a}-Dur{\'a}n, Pablo and Font, Jos{\'e} A. and Yazadjiev, Stoytcho S.",
    title = "{Descalarization bursts from Galactic supramassive neutron stars: a kHz continuous-wave target}",
    eprint = "2610.10728",
    archivePrefix = "arXiv",
    primaryClass = "astro-ph.HE",
    month = "10",
    year = "2026"
}

@article{Rosca-Mead:2020bzt,
    author = "Rosca-Mead, Roxana and Moore, Christopher J. and Sperhake, Ulrich and Agathos, Michalis and Gerosa, Davide",
    title = "{Structure of neutron stars in massive scalar-tensor gravity}",
    eprint = "2007.14429",
    archivePrefix = "arXiv",
    primaryClass = "gr-qc",
    doi = "10.3390/sym12091384",
    journal = "Symmetry",
    volume = "12",
    number = "9",
    pages = "1384",
    year = "2020"
}

@article{Tuna:2022qqr,
    author = {Tuna, Semih and {\"U}nl{\"u}t{\"u}rk, K{\i}van{\c{c}} {\.I}. and Ramazano{\u{g}}lu, Fethi M.},
    title = "{Constraining scalar-tensor theories using neutron star mass and radius measurements}",
    eprint = "2204.02138",
    archivePrefix = "arXiv",
    primaryClass = "gr-qc",
    doi = "10.1103/PhysRevD.105.124070",
    journal = "Phys. Rev. D",
    volume = "105",
    number = "12",
    pages = "124070",
    year = "2022"
}

@article{Demirboga:2023ktt,
    author = "Demirbo{\u{g}}a, Ekrem S. and {\c{S}}ahin, Yakup Emre and Ramazano{\u{g}}lu, Fethi M.",
    title = "{Subtleties in constraining gravity theories with mass-radius data}",
    eprint = "2303.01910",
    archivePrefix = "arXiv",
    primaryClass = "gr-qc",
    doi = "10.1103/PhysRevD.108.024028",
    journal = "Phys. Rev. D",
    volume = "108",
    number = "2",
    pages = "024028",
    year = "2023"
}

@article{East:2021bqk,
    author = "East, William E. and Ripley, Justin L.",
    title = "{Dynamics of Spontaneous Black Hole Scalarization and Mergers in Einstein-Scalar-Gauss-Bonnet Gravity}",
    eprint = "2105.08571",
    archivePrefix = "arXiv",
    primaryClass = "gr-qc",
    doi = "10.1103/PhysRevLett.127.101102",
    journal = "Phys. Rev. Lett.",
    volume = "127",
    number = "10",
    pages = "101102",
    year = "2021"
}

@article{Doneva:2023oww,
    author = "Doneva, Daniela D. and Arest{\'e} Sal{\'o}, Llibert and Clough, Katy and Figueras, Pau and Yazadjiev, Stoytcho S.",
    title = "{Testing the limits of scalar-Gauss-Bonnet gravity through nonlinear evolutions of spin-induced scalarization}",
    eprint = "2307.06474",
    archivePrefix = "arXiv",
    primaryClass = "gr-qc",
    doi = "10.1103/PhysRevD.108.084017",
    journal = "Phys. Rev. D",
    volume = "108",
    number = "8",
    pages = "084017",
    year = "2023"
}

@article{R:2022hlf,
    author = "R, Abhishek Hegade K. and Ripley, Justin L. and Yunes, Nicol{\'a}s",
    title = "{Where and why does Einstein-scalar-Gauss-Bonnet theory break down?}",
    eprint = "2211.08477",
    archivePrefix = "arXiv",
    primaryClass = "gr-qc",
    doi = "10.1103/PhysRevD.107.044044",
    journal = "Phys. Rev. D",
    volume = "107",
    number = "4",
    pages = "044044",
    year = "2023"
}

@article{Ramazanoglu:2016kul,
    author = "Ramazano{\u{g}}lu, Fethi M. and Pretorius, Frans",
    title = "{Spontaneous Scalarization with Massive Fields}",
    eprint = "1601.07475",
    archivePrefix = "arXiv",
    primaryClass = "gr-qc",
    doi = "10.1103/PhysRevD.93.064005",
    journal = "Phys. Rev. D",
    volume = "93",
    number = "6",
    pages = "064005",
    year = "2016"
}

\end{document}